\input epsf.tex    

\documentstyle[12pt]{article}

\newcommand{\bmat}{\left(\begin{array}}
\newcommand{\emat}{\end{array}\right)}

\def\yzero{\smash{\hbox{$y\kern-4pt\raise1pt\hbox{${}^\circ$}$}}}

\def\beq{\begin{equation}}
\def\eeq{\end{equation}}
\def\beqa{\begin{eqnarray}}
\def\eeqa{\end{eqnarray}}

\def\-{\hphantom{-}}
\def\ov{\overline}
\def\s2{\frac{1}{2}}

\def\beq{\begin{equation}}
\def\eeq{\end{equation}}
\def\beqa{\begin{eqnarray}}
\def\eeqa{\end{eqnarray}}

\def\IF{\relax{\rm I\kern-.18em F}}
\def\II{\relax{\rm I\kern-.18em I}}

\def\cp{{\cal P}}
\def\IC{\bf C}
\def\IZ{\bf Z}
\def\IR{\bf R}
\def\IS{\bf S}
\def\IP{\bf P}
\def\IX{\bf X}
\def\IY{\bf Y}
\def\ItY{\bf {\tilde Y}}
\def\IW{\bf W}
\def\IM{\bf M}

\def\id{{\bf 1}}

\def\NN{{\cal N}}
\def\Dsl{\,\raise.15ex\hbox{/}\mkern-13.5mu D} 

\def\IT{\bf T}

 \def\cp#1{\relax\ifmmode {\IP\kern-2pt{}_{#1}}\else $\IP\kern-2pt{}_{#1}$\=fi}
\newcommand{\drawsquare}[2]{\hbox{%
\rule{#2pt}{#1pt}\hskip-#2pt
\rule{#1pt}{#2pt}\hskip-#1pt
\rule[#1pt]{#1pt}{#2pt}}\rule[#1pt]{#2pt}{#2pt}\hskip-#2pt
\rule{#2pt}{#1pt}}

\newcommand{\fund}{\raisebox{-.5pt}{\drawsquare{6.5}{0.4}}}
\newcommand{\Yasymm}{\raisebox{-3.5pt}{\drawsquare{6.5}{0.4}}\hskip-6.9pt%
        \raisebox{3pt}{\drawsquare{6.5}{0.4}}}
\newcommand{\antifund}{\overline{\fund}}

\begin{document}

\makeatletter \@addtoreset{equation}{section} \makeatother
\renewcommand{\theequation}{\thesection.\arabic{equation}}
\pagestyle{empty}
\vspace*{.5in}
\rightline{IFT-UAM-CSIC-02-32}
\rightline{\tt hep-th/0208014}
\vspace{2cm}

\begin{center}
\LARGE{\bf Local models for intersecting brane worlds \\[10mm]}
\medskip
\large{Angel M. Uranga \footnote{\tt Angel.Uranga@uam.es} \\[2mm]}
I.M.A.F.F. and Instituto de F\'{\i}sica Te\'orica C-XVI \\
Universidad Aut\'onoma de Madrid, 28049 Madrid, Spain \\ [3mm]

\smallskip

\small{\bf Abstract} \\[3mm]
\end{center}

\begin{center}
\begin{minipage}[h]{14.5cm}
{\small 
We describe the construction of configurations of D6-branes wrapped on 
compact 3-cycles intersecting at points in non-compact Calabi-Yau 
threefolds. Such constructions provide local models of intersecting brane 
worlds, and describe sectors of four-dimensional gauge theories with 
chiral fermions. We present several classes of non-compact manifolds with 
compact 3-cycles intersecting at points, and discuss the rules required 
for model building with wrapped D6-branes. The rules to build 3-cycles are 
simple, and allow easy computation of chiral spectra, RR tadpoles and the 
amount of preserved supersymmetry. We present several explicit examples 
of these constructions, some of which have Standard Model like gauge group 
and three quark-lepton generations. In some cases, mirror symmetry relates 
the models to other constructions used in phenomenological D-brane model 
building, like D-branes at singularities. Some simple $\NN=1$
supersymmetric configurations may lead to relatively tractable $G_2$ 
manifolds upon lift to M-theory, which would be non-compact but 
nevertheless yield four-dimensional chiral gauge field theories.}

\end{minipage}
\end{center}
\newpage
\setcounter{page}{1} \pagestyle{plain}
\renewcommand{\thefootnote}{\arabic{footnote}}
\setcounter{footnote}{0}

\section{Introduction}
\label{intro}

There has been a lot of recent progress in model building of four-dimensional 
chiral string models with D6-branes wrapped on 3-cycles, mostly centered 
on phenomenological models \cite{bgkl,afiru,imr,bkl,rest,susy,bbkl} (see 
\cite{orbif} for earlier work, and \cite{kachru} for theoretical issues in 
mainly non-chiral models).

All such models are compact, hence to define the gauge sector one needs to 
specify the full model, and to some extent one would have to deal with 
difficult gravitational issues (for instance the issue of the 
cosmological constant in non-supersymmetric models). Moreover, most models 
have been worked out in the particular case of six-torus (or 
orbifolds/orientifolds thereof), with only partial success in extending to 
Calabi-Yau models \cite{bbkl}. This makes the generality of the results 
questionable.

On the other hand, other setups like D3-branes located at singularities  
allowed a more interesting bottom-up approach, where one cooks up a local 
model (D-branes at a singular point in a non-compact Calabi-Yau) which 
might subsequently be regarded as a local description for a small region 
in a full-fledged compactification with four-dimensional gravity. This 
allows to decouple the gauge sector (which mainly depends on the local 
structure) from the details of the choice of compactification manifold, 
and from many difficult (although clearly important) gravitational issues.

\begin{figure}
\begin{center}
\centering
\epsfysize=4.5cm
\leavevmode
\epsfbox{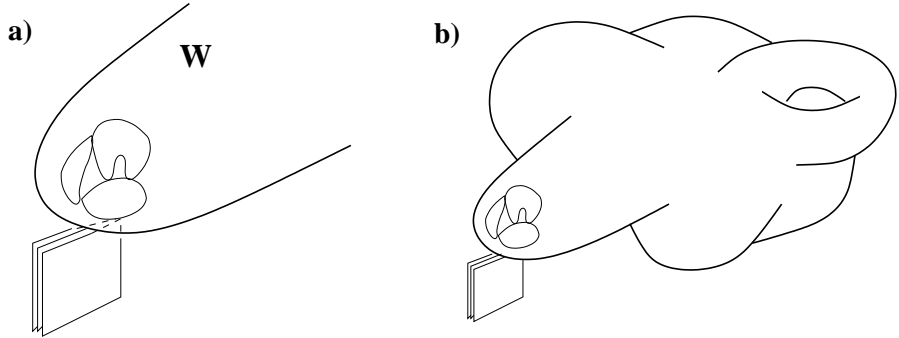}
\end{center}
\caption[]{\small Bottom-up approach to embedding the gauge theory sectors 
from intersecting D-branes in string compactification. In a first step a) 
one considers local configurations of D6-branes wrapped on compact 
3-cycles in a non-compact space $\IW$. The local configuration may be
subsequently embedded in a global compactification, step b). Many features 
of the chiral gauge field theory on the D6-branes are not sensitive to the 
details of the compactification and depend only on the local geometry 
$\IW$.}
\label{bottomup}
\end{figure}

It would be highly desirable to have a similar bottom-up approach for 
models of intersecting D6-branes. This is the task that we undertake in 
the present paper. Namely we are interested in non-compact Calabi-Yau 
spaces with compact 3-cycles which intersect at points. D6-branes wrapped 
on such 3-cycles lead to four-dimensional gauge field theories with chiral 
fermions arising at intersections, just as in toroidal models. In 
addition, the models may contain D6-branes wrapped in non-compact 
3-cycles; these represent global symmetries from the viewpoint of the 
four-dimensional theory (their gauge coupling is infinitely suppressed due 
to the infinite volume) and may become gauged in a fully compact model 
(depending on the details of the compactification). Such local models may 
subsequently be embedded in a full-fledged global compactification, with 
four-dimensional gravity. The important point is that many features of the 
local D-brane configuration, and hence of the gauge sector of the theory, 
are insensitive to the details of the compactification. The idea is 
illustrated in figure \ref{bottomup}

In the present paper we present several classes of non-compact Calabi-Yau 
threefolds with compact 3-cycles intersecting at points. The fact that 
such local Calabi-Yau spaces exist is clear, for instance from mirror 
symmetry as follows. As discussed above, D3-branes at singularities lead 
to localized four-dimensional gauge theories with chiral fermions. From the 
viewpoint of D-branes on Calabi-Yau spaces, systems of D3-/D7- branes at 
singularities correspond to the so-called B type branes, namely can be 
regarded as higher dimensional D-branes wrapped on holomorphic cycles of the 
Calabi-Yau $\IM$ and carrying holomorphic gauge bundles \cite{ooy}. Upon 
application of mirror symmetry, they are mapped to A type branes, in this 
case D6-branes wrapped on special lagrangian 3-cycles in the mirror geometry 
$\IW$ \cite{ooy}, which is non-compact as well
\footnote{A particular case of mirror symmetry (in a global rather than 
local context) is the T-duality between toroidal models with D9-branes 
with magnetic world-volume fluxes (see \cite{bachas,bgkl,magnetised} for 
recent constructions) and toroidal models with D6-branes on 3-cycles; 
consistently with \cite{syz}.}. Since mirror symmetry relates 
configurations yielding the same physics, the final 
D6-branes in $\IW$ must wrap compact 3-cycles intersecting at points, so 
as 
to reproduce the four-dimensional gauge theories with chiral fermions 
obtained from $\IM$.

In fact, several interesting Calabi-Yau threefolds $\IW$ are familiar from 
the mirror symmetry literature. However, even though mirror symmetry 
is a fascinating subject, in this paper we take the viewpoint of 
studying interesting non-compact threefolds with potential to lead to 
interesting four-dimensional chiral gauge theories by wrapping D6-branes 
on compact 3-cycles. We are interested in providing a set of rules to 
compute their spectra for diverse Calabi-Yau constructions, regardless of 
whether or not their mirror manifolds $\IM$ are known \footnote{However, 
we 
would like to point out that some of our results (like for instance the 
description of D6-branes mirror to D7-branes at singularities, section 
\ref{standard}) are not described in the mirror symmetry literature, hence 
should interest the corresponding readers.}.

We will describe the basic rules for the construction of compact 
intersecting 3-cycles in two large classes of non-compact Calabi-Yau 
spaces. Within each class we will present simple examples of models with 
spectrum similar to the Standard Model, simply to illustrate the richness 
of the constructions. More detailed search for phenomenological models is 
left for future work.

One clear advantage of the use of local configurations in model building 
is that they allow a better handle on the final spectrum of the gauge 
sector of the theory. A second advantage, from the phenomenological 
viewpoint, is that models with D6-branes wrapped on 3-cycles localized 
in a small region of the Calabi-Yau allow for a low string scale. Indeed, 
by enlarging the dimension of the Calabi-Yau which are transverse to the 
set of D6-branes one may reproduce a large four-dimensional Planck mass, 
even for low string scales, without enlarging the size of 3-cycles 
(therefore without the risk of generating too light Kaluza-Klein replicas 
of Standard Model gauge bosons). This possibility was not allowed in 
toroidal-like models, since they contain no direction which is transverse 
to all D6-branes \cite{bgkl}. 

Finally, some of the wrapped 3-cycles in our models are topologically 
3-spheres, hence the corresponding gauge groups do not contain adjoint 
$\NN=1$ chiral multiplets. This is in contrast with toroidal models (and 
quotients thereof) where the 3-cycles used had non-trivial $b_1$ and lead 
to a number of (phenomenologically undesirable) adjoint matter multiplets.

The paper is organized as follows. In Section \ref{sthreefold} we describe 
a class of non-compact Calabi-Yau threefolds, describe several kinds of 
compact and non-compact 3-cycles in them, and construct explicit examples 
of local intersecting brane worlds with D6-branes wrapped on 3-cycles. Some 
of them have gauge sector spectrum close to the Standard Model. In Section 
\ref{generalizations} we briefly sketch several generalizations of the 
above construction in more involved threefolds, with a richer structure 
of compact 3-cycles. Our treatment aims at stating the basic model 
building rules, and we leave their more systematic exploration for future 
work. In Section \ref{twofold} we describe manifolds given by a product 
of a two-torus times a non-compact two-fold. We describe diverse 3-cycles 
in such space, and use them to build chiral gauge sectors of intersecting 
D6-branes. In Section \ref{imr} we discuss the construction of some  
explicit configurations, in certain (non-CY) threefold with topology 
$\IT^2\times \IT^2\times \IY$, with $\IY$ a non-compact Riemann surface,
and producing {\em just} the Standard Model gauge interactions and chiral 
fermion content. In Section \ref{conclusions} we present our concluding 
remarks.

\section{Local intersecting brane worlds in non-compact threefolds}
\label{sthreefold}

In this section we describe a class of non-compact threefolds with compact 
3-cycles intersecting at points. They have been studied in 
\cite{hiv,hi,cfikv,fhhi} in a different context (building on earlier work 
\cite{timoy}), so we find it 
useful to review their basic results.

\subsection{The threefolds}

\begin{figure}
\begin{center}
\centering
\epsfysize=4.5cm
\leavevmode
\epsfbox{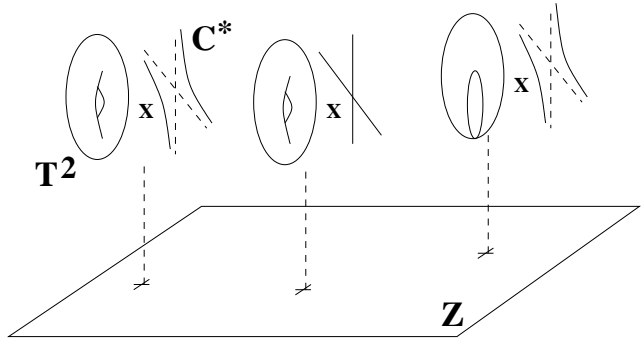}
\end{center}
\caption[]{\small Schematic depiction of the fibration structure of the 
threefold $\IW$. At certain points on the $z$-plane, the $\IC^*$ or the 
elliptic fiber degenerate.}
\label{threefold}
\end{figure}

Let us consider the following class of non-compact Calabi-Yau threefolds, 
given by the equations
\beqa
u\, v\, & = & z \nonumber \\
y^2 \, & = & x^3 \, + \, f(z)\, x \, + \, g(z)
\label{wthreefold}
\eeqa
It corresponds to a non-compact complex plane, parametrized by $z$, over 
which we have a $\IC^*$ fibration 
times an elliptic fibration. The structure of the fibration is sketched in 
figure \ref{threefold}. Over a generic point $z_0$, the $\IC^*$ fiber
is given by $uv=z_0$, and contains an $\IS^1$ (e.g. for real $z_0$, the 
$\IS^1$ is made manifest by introducing $u=x_1+ix_2$, $v=x_1-ix_2$ and 
taking $x_i$ real in $x_1^2+x_2^2=z_0$). At $z=0$ the $\IC^*$ fiber 
degenerates to two complex planes, and the $\S^1$ shrinks to zero. 

Over a generic point the fiber of the elliptic fibration 
is a two-torus; the fiber degenerates at points $z_a$ which we will label
by the $(p,q)$ 1-cycle that collapses. The degeneration points and their 
$(p,q)$ nature depend on the polynomials $f(z)$, $g(z)$ and we specify 
them in the discussions below. The 1-cycles in the fiber of the elliptic 
fibration suffer $SL(2,Z)$ monodromies in going around degenerate fibers. 
Hence, out of each degeneration point of the elliptic fibration, there is 
a branch cut in the $z$ plane going off to infinity. In crossing 
(counterclockwise) a branch cut due to a $(p_a,q_a)$ degeneration, an 
$(r,s)$ 1-cycle suffers a monodromy and turns into an $(r',s')$ one, with
\beqa
\pmatrix{r \cr s} \rightarrow \pmatrix{r'\cr s'} \, =\, \pmatrix{
1+p_aq_a & -p_a^2 \cr q_a^2 & 1-p_aq_a} \cdot \pmatrix{r \cr s}\, = \, 
\pmatrix{r\cr s}\, +\, (rq_a-s p_a)\, \pmatrix{p_a\cr q_a}
\label{monodromy}
\eeqa
The location of these branch cuts is unphysical. However, the $(p,q)$ 
labels of degenerations themselves may change by moving the branch cuts. 
Therefore it is convenient to fix a canonical representation for the 
elliptic fibration; without loss of generality in the 
discussion of topological properties \footnote{Beyond topology, the 
precise location and nature of the $(p,q)$ degenerations varies over
moduli space, or more precisely, over a suitable cover of moduli space 
(Teichm\"uller space).}, we choose to order the degenerations 
in counterclockwise circle around $z=0$ in the $z$-plane with branch cuts 
emanating outwards. We will denote such configuration by 
\beqa
(p_1,q_1), (p_2,q_2), \ldots, (p_N,q_N)
\eeqa

We will not need to move degeneration points through branch cuts
\footnote{A beautiful connection, which we will not exploit in the present 
paper, relates Picard-Lefschetz transformations (transformations where 
degeneration points move around branch cuts and return to their original 
positions) to non-trivial transformation on field theory on D-brane probes 
(Seiberg dualities). See \cite{oovafa,ito,fhhu,cfikv,fiol,fhhi,fh,bd} for 
further discussions.}; however for completeness we point out that the 
rules for such crossings are
\beqa
 (p_i,q_i)\, ,\, (p_{i+1}, q_{i+1}) & \rightarrow & (p_{i+1},q_{i+1})\, , 
\, [\, (p_i,q_i)+I_{i,i+1}(p_{i+1},q_{i+1})\, ] \nonumber \\
 (p_i,q_i)\, ,\, (p_{i+1}, q_{i+1}) & \rightarrow & 
[\, (p_{i+1},q_{i+1})+I_{i,i+1}(p_{i},q_{i})\, ] \, , \,  (p_i,q_i)
\eeqa
with  $I_{i,i+1}=p_iq_{i+1}-q_ip_{i+1}$.

We will shortly see that this class of threefolds are interesting in 
that, although they are non-compact still contain a rich structure of 
compact 3-cycles. Threefolds of this kind (with degenerations specified 
below) are familiar from the literature on mirror symmetry, since they
provide the (local) mirrors of Calabi-Yau threefold singularities given by 
complex cones over del Pezzo surfaces \footnote{For the interested reader, 
the total space is the anticanonical bundle over the del Pezzo surface. 
The $k^{th}$ del Pezzo surfaces can be constructed as the blow up of 
$\IP_2$ at $k$ generic points or the blow up of $\IP_1\times\IP_1$ at 
$k-1$ points ($0\leq k \leq 8$).}. The set of $N$ degenerate elliptic 
fibers in the mirror $\IW$ of the $N^{th}$ del Pezzo surface is given (in 
canonical representation) by
\beqa 
(1,0),\ldots,(1,0), (2,-1), (-1,2), (-1,-1) 
\label{dege} 
\eeqa 
It is interesting to point out that the complex cone over $\IP_2$ is the
familiar $\IC^3/\IZ_3$ orbifold singularity; we will use its mirror, which
is the manifold (\ref{wthreefold}) with degenerated elliptic fibers
$(2,-1), (-1,2), (-1,-1)$, in many of our explicit discussions below.  
However, extension to other threefolds should be clear.

For future convenience we note that the total monodromy due to the branch 
cuts of the above set (\ref{dege}) of degenerate elliptic fibers is 
$T^{12-N}$ with 
\beqa
T\, =\, \pmatrix{1 & 1 \cr 0 & 1}
\label{tconjugate}
\eeqa
which leaves the 1-cycle $(1,0)$ invariant. This total monodromy will turn 
out important in the construction of certain compact 3-cycles, see 
Section~\ref{trinification}. However, it is not necessary for consistency, 
and other elliptic fibration with other sets of degenerate fibers can be 
studied using similar techniques. In \cite{sz} the authors defined and 
classified {\em isolated} configurations of $(p,q)$ degenerations of 
elliptic fibers, namely those that may be obtained from compact elliptic 
fibrations by taking a non-compact limit. In order to keep within the 
non-compact setup one should restrict to those configurations
\footnote{Even though the techniques in our paper are valid to construct 
3-cycles in compact fibrations as well, additional constraints from 
global tadpole cancellation would complicate the analysis beyond our main 
interest.}. Any elliptic fibration of this kind would provide a sensible 
threefold \footnote{One may be willing to relax the Calabi-Yau condition 
and consider non-supersymmetric threefolds; to our knowledge there is no 
easy rule to determine which sets of degenerations give sensible 
manifolds. One may always start with the Weierstrass form for the elliptic 
fibration and determine the monodromies due to degenerate fibers.}.
Even though we will mainly center on the set of degenerations (\ref{dege})
for examples, our techniques apply to other examples.

We would like to emphasize that even though these threefolds first 
appeared in the mirror symmetry context, we are not particularly 
interested in mirror symmetry. Rather, we use these familiar threefolds to 
build models of local configurations of D6-branes wrapped on intersecting 
3-cycles. We expect the results and techniques learnt in this context to 
extend to other threefolds whose mirror symmetry properties are not 
particularly interesting/known. However we will make some side remarks on 
the mirror symmetry interpretation of some of our D6-branes along the way.

\subsection{D6-branes on 3-cycles}
\label{threecycles}

\subsubsection{A simple set of compact 3-cycles}

Let us consider a general threefold $\IW$ of the above kind, with the 
elliptic fibration specified by a set of $(p_a,q_a)$ degeneration points
in canonical representation. We now describe a simple class of compact 
3-cycles in this non-compact space, at the topological level. See section 
\ref{supersymmetry} for a discussion of supersymmetry conditions.

\begin{figure}
\begin{center}
\centering
\epsfysize=3cm
\leavevmode
\epsfbox{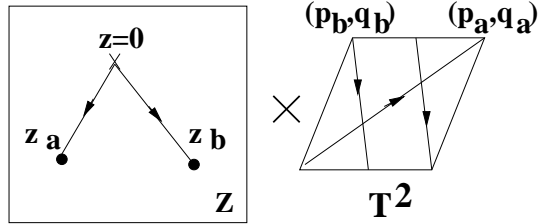}
\end{center}
\caption[]{\small Schematic depiction of the intersection between two 
compact 3-cycles. The two 3-cycles touch at $z=0$ in the $z$-plane, where 
they also touch in the $\IC^*$ fiber since the $\IS^1$ shrinks to zero. On 
the elliptic fiber, the 3-cycles intersect according to their $(p,q)$ 
labels.}
\label{intersection}
\end{figure}

Consider a segment in the $z$-plane that joins (without crossing any 
branch cut) $z=0$ with the $(p_a,q_a)$ degeneration point $z_a$. A 
compact 3-cycle with no boundary is obtained by taking this segment and 
fibering over it the $\IS^1$ in the $\IC^*$ fibration times the 
$(p_a,q_a)$ 1-cycle in the elliptic fibration, see figures 
\ref{intersection}, \ref{junction}a. We will denote such cycle a 
$(p_a,q_a)$ 3-cycle.
This 3-cycle is topologically a 3-sphere, as follows. Near the pinch of a 
degenerate fiber at $z_0$, an elliptic fibration may be described by 
$u'v'=z-z_0$. Taking this equation along with the $\IC^*$ fibration 
$uv=z$, we obtain $uv-u'v'=z_0$. By changing variables we can recast this 
as
\beqa
x_1^2\, +\, x_2^2 \, +\, x_3^2\, +\, x_4^2\, =\, z_0
\eeqa
Taking e.g. $z_0$ real, our 3-cycle is given by taking real $x_i$, which 
describes an $\IS^3$ with size controlled by $z_0$, the length of the 
segment on the $z$ plane. 

Two such $(p_a,q_a)$ and $(p_b,q_b)$ 3-cycles generically intersect at 
several points, located on the fiber over $z=0$ (where they coincide both 
in the $z$-plane and in the $\IC^*$ fibration), see figure 
\ref{intersection}. Their intersection number when both are identically 
oriented in $\IC^*$ and on the $z$-plane (say, both outgoing from $z=0$) 
is given by
\beqa
I_{ab}\, = \, p_a \, q_b \, - \, q_a\, p_b
\eeqa
Hence D6-branes wrapped on these 3-cycles are interesting examples of 
local configurations of intersecting D6-branes giving  four-dimensional 
gauge field theories with chiral fermions. Explicit examples are discussed 
in Section~\ref{examples}.

This set of 3-cyles is moreover important in that it provides a basis for 
the compact 3-homology of the manifold $\IW$. Hence any other compact 
3-cycle can be expressed topologically as a linear combination of 3-cycles
in the above set. However, the minimal volume 3-cycle representative of a 
given homology class may not be a 3-cycle of the above simple kind, as we 
describe in the following.

\subsubsection{3-cycles from junctions of $(p,q)$ segments}

\begin{figure}
\begin{center}
\centering
\epsfysize=4.5cm
\leavevmode
\epsfbox{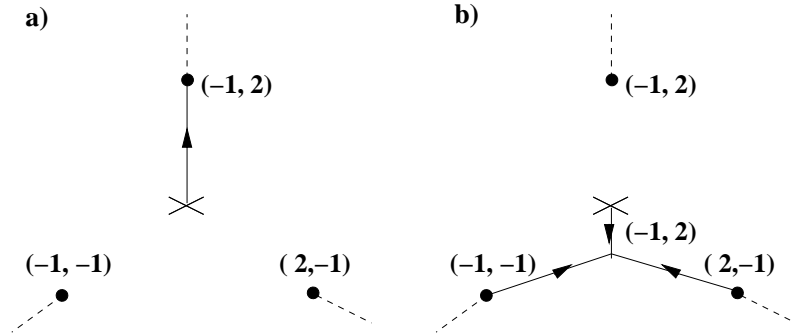}
\end{center}
\caption[]{\small Two examples of different compact 3-cycles. Fig a) shows 
a compact 3-cycles obtained by fibering over a segment in the $z$-plane
a $(p,q)=(-1,2)$ 1-cycle in the elliptic fiber and the $\IS^1$ in the 
$\IC^*$ fibration. Fig b) shows a compact 3-cycle obtained by fibering 
suitable $(p,q)$ 1-cycles over segments over a network in the $z$-plane. 
Note the conservation of $(p,q)$ wrapping numbers across the junction.}
\label{junction}
\end{figure}

There are more general compact 3-cycles that the above ones. They can be 
obtained by taking a network of oriented segments in the $z$-plane, 
joining at junctions, and with outer legs ending on the $(p_a,q_a)$ 
degenerations or at $z=0$. Over such segments we fiber the $\IS_1$ in the 
$\IC^*$ fibration and a $(p,q)$ 1-cycle in the elliptic fibration. To 
obtain a closed 3-cycle the $(p,q)$ label of the segments must be 
additively conserved at each junction, and segments ending on a 
degeneration of the elliptic fibration must be of the correct $(p,q)$ 
kind. One example of a compact 3-cycle from a junction of $(p,q)$ segments 
is shown in figure \ref{junction}.

The rules concerning 3-cycles from junctions are identical to the rules 
for junctions of $(p,q)$ strings in type IIB seven-brane backgrounds 
\footnote{These rules have also appeared in the study of networks of type 
IIB five-branes in seven-brane backgrounds \cite{dhik}.}. In fact our 
configurations are related by U-duality to type IIB configurations with a 
set of $(p_a,q_a)$ seven-branes (U-dual to our degenerations of the 
elliptic fibration), one D3-brane (U-dual to the degeneration of the 
$\IC^*$ fibration), and $(p,q)$ strings (U-dual to the D6-branes on 
3-cycles) \footnote{The duality chain proceeds as follows. We start with a 
space roughly of the form $\IC^*\times \IT^2\times \IC_{z}$, with fibered 
products, with one $\IC^*$ degeneration, a set of $(p,q)$ elliptic fiber 
degenerations, and D6-branes on 3-cycles made of $(r,s)$ segments. 
T-dualizing along the three spatial Minkowski directions turns the 
D6-branes into D3-branes on the same 3-cycle. Further T-duality along the 
$\IS^1$ direction in $\IC^*$ turns the $\IC^*$ degeneration into a 
NS5-brane, and the D3-branes into D2-branes. Subsequent lift to 
M-theory leads to a geometry $\IR\times \IS^1\times \IT^2\times \IC_{z}$ 
with one M5-brane, a set of $(p,q)$ degenerate $\IT^2$ fibers, and 
M2-branes on $(r,s)$ 3-cycles. Upon shrinking $\IT^2$ one recovers a IIB 
model with one D3-brane, a set of $(p,q)$ seven-branes and a network of 
$(r,s)$ strings.}. The topological properties of string junctions have 
been extensively studied 
(see e.g. \cite{ghz,oz,dewolfe,dhizkod,dhizkm,hiz,sz}) and we have 
found those results useful in studying 3-cycles. In fact, some of the 
above (and below) conventions and phenomena are taken from standard string 
junction techniques. However we have chosen to briefly re-describe them in 
our present context.

\begin{figure}
\begin{center}
\centering
\epsfysize=3.5cm
\leavevmode
\epsfbox{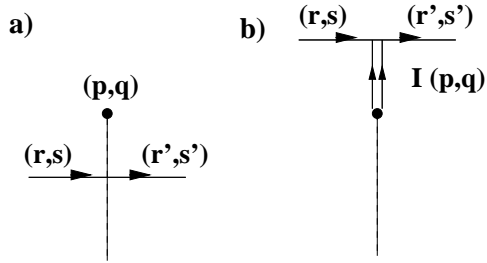}
\end{center}
\caption[]{\small The prong creation process shows the existence of 
junction 3-cycles in our geometries.}
\label{prong1}
\end{figure}    

\begin{figure}
\begin{center}
\centering
\epsfysize=3.5cm
\leavevmode
\epsfbox{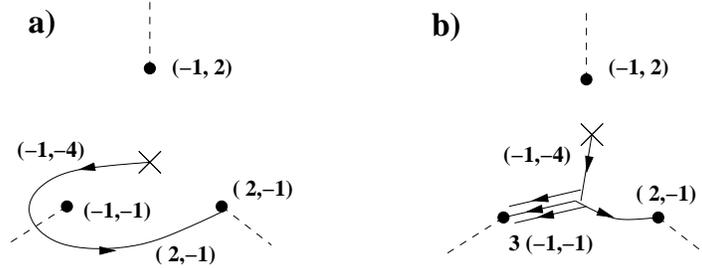}
\end{center}
\caption[]{\small An example of a prong creation process with compact 
3-cycles.}
\label{prong2}
\end{figure}

In analogy with string junctions, it is possible to derive the existence 
of junction 3-cycles by the prong creation process \cite{ghz}, illustrated 
in figures \ref{prong1}, \ref{prong2}. Consider an $(r,s)$ segment crossing 
a $(p,q)$ branch cut, 
hence turning into an $(r',s')$ segment, see (\ref{monodromy}). 
Moving the segment across the $(p,q)$ degeneration point, prongs emanating 
from the latter are created, and the 3-cycle becomes a junction 3-cycle.
The number of prongs outgoing from the degeneration point is given by
\beqa
I_{(r,s),(p,q)}\, =\, rq-ps.
\eeqa
This process can be understood from charge conservation on the volume of 
the $(p,q)$ degeneration, when regarded as an object, or from geometrical 
considerations as in \cite{oovafa}. It is ultimately related by U-duality 
to the Hanany-Witten brane creation process \cite{hw}.

Intersection numbers of junction 3-cycles are easily computed by careful 
addition (with signs) of intersections of different segments. Notice that 
two segments crossing away from $z=0$ give zero contribution since
since do not intersect on the $\IC^*$ fiber.

\subsubsection{Non-compact 3-cycles}

It is clear that we have restricted our discussion to compact 3-cycles 
only for convenience. The setup immediatly allows to construct non-compact 
3-cycles by using networks of $(p,q)$ segments with some legs going off to 
infinity, over which we fiber the $\IS^1$ in the $\IC^*$ fibration times 
the corresponding $(p,q)$ 1-cycle in the elliptic fibration. In order to 
obtain consistent 3-cycles, external segments must end at infinity, at a 
suitable $(p,q)$ degeneration, or at the $\IC^*$ degeneration point $z=0$.

Intersection numbers between compact and non-compact 3-cycles are computed 
straightforwardly. However, intersection numbers between non-compact 
3-cycles require for their definition some information about asymptotic 
behaviour (in other words, there may be some intersection number 
contribution lying at infinity, which would become manifest only upon 
compactification). To avoid complications, we will not discuss 
intersection numbers among non-compact 3-cycles.

\medskip

In this section we would like to introduce a different kind of non-compact 
3-cycles, which, instead of wrapping the $\IS^1$ in the $\IC^*$ fibration,
span the non-compact direction. At a generic point $z_0=re^{i\theta}$, the 
latter is given e.g. by taking $u=s e^{i\theta/2}$, $v=r/s e^{i 
\theta/2}$ in $uv=z_0$, with $s\in \IR$ parametrizing the non-compact 
line. The non-compact 3-cycles we are interested in are obtained by 
taking networks of $(p,q)$ segments (extending to infinity in the 
$z$-plane or not) and fibering the non-compact direction 
in the $\IC^*$ fibration, times the corresponding $(p,q)$ 1-cycle in the 
elliptic fibration. Notice that for these 3-cycles no external segment is 
allowed to end on the $\IC^*$ degeneration point, since this would result 
on a 3-cycle with boundary.

There are two important differences between 3-cycles wrapping the $\IS^1$ 
or the non-compact directions in $\IC^*$ (momentarily denoted $a$ and $b$ 
3-cycles, respectively). The first is topological: A non-compact $a$-cycle 
does not intersect other $a$-cycle (whether compact or not) unless both 
have segments ending on the $\IC^*$ degeneration point. In particular, 
segments crossing away from $z=0$ do not lead to net intersection number, 
since both cycles are parallel on the $\IC^*$ fiber. On the other hand, a 
non-compact $b$-cycle intersects an $a$-cycle at any crossing of segments, 
with intersection number given by $I_{(p,q),(p',q')}$. This makes 
non-compact $b$-cycles more interesting that non-compact $a$-cycles
\footnote{There are also further sources of intersection which we will not 
need, but list here for completeness: segments of the $a$ and $b$ kind 
ending on the same $(p,q)$ degeneration contribute to the intersection 
number. Similarly, a $(r,s)$ segment crossing the branch cut of a $(p,q)$ 
degeneration contributes an intersection number $I_{(r,s),(p,q)}$ with a 
segment ending on the $(p,q)$ degeneration, if they are of $a$ and $b$ kind, 
or viceversa. This can be seen by moving the $(r,s)$ segment through the 
degeneration; the newly created prongs do intersect the $(p,q)$ segment.}.

The second important difference concerns the conditions for supersymmetry, 
so we postpone its discussion until next section. 

A last important fact concerning 3-cycles of $b$ kind is the following: 
when such a 3-cycle with label $(p,q)$ crosses over $z=0$, a prong 
emanating from $z=0$ (with suitable orientation) is created, corresponding 
to a piece of $a$-cycle 
with $(p,q)$ label. Creation of this prong is due to the fact that the 
non-compact direction in $\IC^*$ intersects the 1-cycle degeneration at 
$z=0$ (namely, the $\IS^1$ in $\IC^*$), and can be shown as previous 
prong creations processes  \footnote{For instance by recalling that a 
$\IC^*$ fibration near a degeneration can be constructed from an elliptic 
fibration near a degeneration by taking the decompactification limit of 
the fiber. The $a$ and $b$ cycles in the torus turn into the $a$ and $b$ 
cycles in the $\IC^*$}. This fact will be useful in some deformation 
arguments in section \ref{standard}.

\subsection{Supersymmetry and calibrations}
\label{supersymmetry}

Different D-branes of A type (e.g. D6-branes wrapped on 3-cycles) preserve 
a common supersymmetry if they 
calibrate with respect to the same 3-form $\Omega_3$. Namely, if they are 
special lagrangian (slag) with respect to $\Omega_3$. A 3-cyles is special 
lagrangian if the restriction of the K\"ahler form $J$ vanishes, and if 
the imaginary part of the restriction of the calibrating 3-form vanishes
($Im(\Omega_3)=0$ restricted to the 3-cycle). For the 3-cycles that we 
consider the former condition is automatically satisfied, hence we 
just need to impose the latter.

For flat geometries like tori and factorized 3-cycles, the supersymmetry 
conditions can be stated in terms of simple condition of the angles of the 
3-cycle in the different complex planes. Even though our geometries are 
more involved, it is remarkable that the supersymmetry conditions can be 
similarly stated as simple conditions on the orientation of the 3-cycles 
on the $z$-plane, the $\IC^*$, and the elliptic fiber.

The geometries at hand are 
\beqa 
y^2\, & = & \, x^3\, +\, f(z)\, x\, +\, g(z) \nonumber \\ 
u\, v\, & = & \, z 
\label{mirror} 
\eeqa 
We may recast
the elliptic fibration in terms of two real variables $x_1$, $x_2$ with
identifications $x_1+ix_2=(x_1+ix_2)+1$ and $x_1+ix_2=(x_1+ix_2)+\tau$
with $\tau(z)$ the fiber torus period \footnote{The period is determined
from (\ref{mirror}) via the familiar modular invariant j-function,
$j(\tau)=\frac{4(24f)^3}{4f^3+27g^2}$}. 
Then a calibrating 3-form is 
\beqa 
\Omega_3 \, = \, i \, \frac{du}{u} \, \wedge \, (dx_1 + i dx_2) \, \wedge 
\, dz
\eeqa

A typical kind of 3-cycles found above is given by
\beqa
u \, & = \, & a \, e^{i\theta} \nonumber \\
v \, & = \, & a \, e^{-i\theta} e^{i\theta_1} \nonumber \\
z \, & = \, & a^2 \, e^{i\theta_1} \nonumber \\
x_1+ix_2 & = & (\, p\, +\, \tau\,  q\, )\, b
\label{threecycleone}
\eeqa
parametrized by $a,b,\theta\in \IR$. It wraps the $\IS^1$ fiber in 
the $\IC^*$ fibration, a real line in the $z$-plane, and the $(p,q)$ 
1-cycle in the elliptic fiber. Denoting $p+\tau q=L e^{i\theta_2}$, 
we have that $\Omega_3$ when restricted to the cycle
\beqa
\Omega_3 \, = \, -2\, L\,  a\, e^{i(\theta_1+\theta_2)} \, d\theta \wedge 
db \wedge da
\eeqa
The 3-cycles is Slag with respect to the 3-form if $\theta_1+\theta_2=0$ 
(if $\theta_1+\theta_2=\pi$ the wrapped D-brane preserves the 
opposite supersymmetries, i.e. behaves as the corresponding antibrane).
With a suitable choice of conventions, this amounts to requiring that the 
slope of the base segment in the $z$-plane is adjusted to its $(p,q)$ 
label.

A second kind of 3-cycle is given by
\beqa
u \, & = \, & a \, e^{i\varphi_0} \nonumber \\
v \, & = \, & b \, e^{-i\varphi_0} e^{i\theta_1} \nonumber \\
z \, & = \, & a\, b \, e^{i\theta_1} \nonumber \\
x_1+ix_2 & = & (\, p\, +\, \tau\,  q\, )\, c
\label{threecycletwo}
\eeqa
paremetrized by $a,b,c\in \IR$. It spans the non-compact direction in 
$\IC^*$, a line in the $z$-plane, and wraps the 
$(p,q)$ 1-cycle in the elliptic fiber. The 3-form restricted to the 
3-cycle gives
\beqa
\Omega_3 \, =\,  i\, L\, e^{i(\pi/2+\theta_1+\theta_2)} \, da\wedge 
db\wedge dc
\eeqa
The 3-cycle is supersymmetric with respect to that 3-form if 
$\theta_1+\theta_2+\pi/2=0$. Notice that the 3-cycle preserves the same 
susy as the previous one (\ref{threecycleone}) if they intersect at angles 
in $SU(3)$ (taking into account the $\pi/2$ angle in $\IC^*$) as in 
\cite{bdl}.

In general we will be happy if every 3-cycle is Slag with respect to {\em 
some} calibrating 3-form. This simply means that the 3-cycle is the 
product of three minimal volume 1-cycles. In particular, they fiber over 
geodesics (straight lines in the pictures) in the $z$-plane. If all 
3-cycles happen to calibrate with a common 3-form (namely their slopes in 
the $z$-plane are adjusted to their $(p,q)$ labels), an overall $\NN=1$ 
supersymmetry will be preserved. If not, intersections are typically 
non-supersymmetric; we expect that the corresponding potential open string
tachyons may be avoided in some regions of moduli space. 

In talking about supersymmetric cycles, an important notion is their 
moduli space. Let us simply quote the standard result \cite{moduli} that for 
a Slag 3-cycle $\Pi$ the (complex) dimension of its moduli space is 
$b_1(\Pi)$. From the field theory viewpoint, this is the number of adjoint 
matter multiplets of the corresponding gauge group. Hence, Slags which 
are topologically $\IS^3$ lead to no adjoint matter, while Slags which are 
topologically $\IT^3$ have three adjoint chiral multiplets.

We conclude by mentioning that in most arguments above and to follow, 
involving only topological properties, we will not care about susy at 
all, even for a single set of branes. Hence we drop the Slag condition 
for topological issues.

\subsection{Tadpole cancelation}

In a compact threefold $\IW$, the RR tadpole cancellation conditions 
for a set of $N_a$ D6$_a$-branes wrapped on 3-cycles with homology classes 
$[\Pi_a]$ are obtained from consistency of the equations of motion of the 
RR 7-form $C_7$. From the action
\beqa
S_{C_7} \, = \, \int_{M_4\times W} H_8\, \wedge\, *H_8 + \sum_a \, N_a 
\, \int_{M_4\times \Pi_a} C_7
\eeqa
(where $H_8$ is the field strength tensor for $C_7$) the equations of 
motion read
\beqa
dH_8 \, =\, \sum_a\, N_a\, \delta(\Pi_a)
\eeqa
with $\delta(\Pi)$ is a bump 3-form with support on the 3-cycle $\Pi$. 
This equation in homology gives an integrability condition (Gauss' law)
\beqa
[\Pi_{\rm tot}]\, =\, \sum_a\, N_a\, [\Pi_a]\, =\, 0
\eeqa
In a non-compact space, RR flux may escape to infinity, so the total 
homology class need not vanish. Rather, charge associated to homology 
classes which may be taken away to infinity by deformation should be 
allowed. Such homology classes correspond to cycles which do not 
intersect any localized 3-cycle (and hence are not in principle topologically 
obstructed to be carried away to infinity, at least for sufficiently 
non-compact manifolds \cite{abiu}). Therefore the RR tadpole conditions 
for non-compact threefolds read
\beqa
[\Pi_{\rm tot}]\cdot [\Sigma_i] \, = \, 0
\label{localrr}
\eeqa
where $[\Sigma_i]$ are a basis of compact 3-cycles. This automatically 
guarantees cancellation of cubic non-abelian anomalies \footnote{
It also guarantees that mixed $U(1)$ - non-abelian anomalies cancel by a 
Green-Schwarz mechanism, involving scalars arising from integrals of the 
RR 3-form over 3-cycles. Its derivation is given in \cite{afiru}, section 
3.2.2, where it was already announced that its validity was beyond the 
toroidal setup.}, as can be easily checked.

Since a general discussion of the contribution of different sets of branes 
to the tadpoles is involved, we postpone the detailed computation of 
tadpoles to the explicit examples below. For the time being, it is enough 
to state that the conditions boil down to requiring cancellation of gauge 
anomalies in the generic gauge theory from D6-branes wrapped on the basic 
compact 3-cycles \footnote{By this we mean requiring the number of 
fundamentals and antifundamentals to be equal for all nodes in the quiver 
gauge theory, regardless of the number of branes associated to such node.}
(as expected from other discussions in non-compact setups where RR 
tadpoles and anomalies were shown to be equivalent \cite{abiu}). Notice 
that in embedding the local model into a global compact one, the remaining 
tadpole may be saturated by adding branes in the correct homology class. 
From the viewpoint of the gauge group in the non-compact case, this would 
be a hidden non-intersecting brane, leaving the local gauge sector 
unchanged (although in the compact setup the brane may become visible, 
if it is charged under groups of previously non-compact branes, since the 
latter become gauged upon compactification).

\subsection{Some examples of model building}
\label{examples}

In this section we describe some simple examples of four-dimensional 
chiral gauge theories which can be obtained using the above threefolds.

\subsubsection{A simple example with matter triplication}
\label{trinification}

\begin{figure}
\begin{center}
\centering
\epsfysize=3.5cm
\leavevmode
\epsfbox{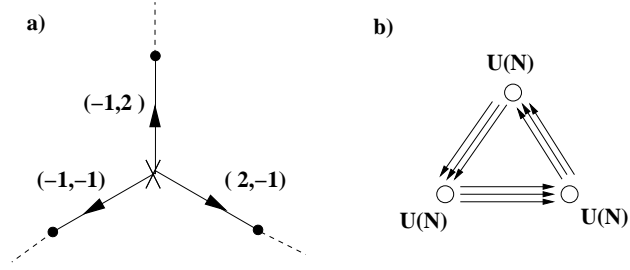}
\end{center}
\caption[]{\small Configuration of D6-branes wrapped on compact 3-cycles 
in a RR tadpole free fashion.}
\label{dthree}
\end{figure}    

Consider the threefold $\IW$ defined by (\ref{wthreefold}) with elliptic 
fiber degenerations
\beqa
(2,-1)\, , \, (-1,2)\, , \, (-1,-1)
\eeqa
Consider a configuration of $N$ D6-branes wrapping each of the three 
$\IS^3$ basic 3-cycles obtained from segments stretching from $z=0$ to 
the $(p,q)$ degeneration points of the elliptic fibration, figure 
\ref{dthree}a. 

\begin{figure}
\begin{center}
\centering
\epsfysize=3.5cm
\leavevmode
\epsfbox{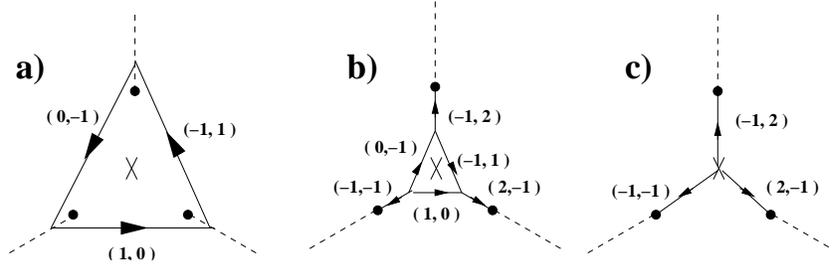}
\end{center}
\caption[]{\small Continuous process showing the total configuration of 
D6-branes in figure \ref{dthree} is homologous to a cycle which can be 
taken to infinity.}
\label{regfract}
\end{figure}

Let us denote $[\Sigma_{(p,q)}]$ the homology class of the corresponding 
3-cycle.
It is easy to check that this configuration satisfies cancellation of RR 
tadpoles. Specifically, the sum of the three basic 3-cycles is homologous 
to a 3-cycle which can be taken away to infinity. The latter 3-cycle is 
given topologically by a closed path around the origin in the $z$-plane 
over which we fiber the $\IS^1$ in the $\IC^*$ fibration times a 1-cycle 
in the elliptic fibration, with $(p,q)$ labels as in figure 
\ref{regfract}a. Note that the existence of this 3-cycle is possible from 
the fact that the total $SL(2,Z)$ monodromy in the elliptic fibration is 
$T^9$, see discussion around (\ref{tconjugate}). The 3-cycle is 
topologically a 3-torus, so it has a three complex dimensional moduli 
space (corresponding to motion in the three transverse directions, plus 
turning on Wilson lines).

Figures \ref{regfract}a, b, c illustrate the continuous process, involving 
brane creation, which connects the homological sum of the three $\IS^3$'s 
to the final 3-cycle, thus proving RR tadpole cancellation in the setup.
Hence, we can write the equation
\beqa
[\Sigma_{(-1,-1)}]\, +\, [\Sigma_{(2,-1)}]\, +\, [\Sigma_{(-1,2)}]\, =\, 0
\eeqa
modulo a homology element without intersections with the compact 3-cycles 
(namely, the homology class of the $\IT^3$ 3-cycle). 

It is interesting to note that for a suitable choice of location of the 
degeneration points, the above configurations are $\NN=1$ supersymmetric, 
i.e. satisfy the Slag condition in section \ref{supersymmetry} for a 
common calibrating form. Namely, the slopes of different segments in the 
$z$-plane are adjusted to their $(p,q)$ labels.
As usual, deformations of complex structure away from that point in 
moduli space result in the generation of Fayet-Iliopoulos terms, 
potentially triggering brane recombination and gauge symmetry breaking.
For simplicity we choose the supersymmetric configuration, as shown in the 
figures.

The intersection numbers are easily computed from our rules
\beqa
[\Sigma_{(-1,-1)}] \cdot [\Sigma_{(2,-1)}]\, = \,
[\Sigma_{(2,-1)}] \cdot [\Sigma_{(-1,2)}]\, = \,
[\Sigma_{(-1,2)}] \cdot [\Sigma_{(-1,-1)}]\, =\, 3
\eeqa

The spectrum of the four-dimensional theory on the D6-branes is
\beqa
{\rm Gauge}\, {\rm  group} \quad & \quad \quad \quad U(N)\times U(N)\times 
U(N) \nonumber \\
\NN=1 \, {\rm Ch.} \,  {\rm Mult.} \quad & \quad 3\,(\fund,\antifund,1) 
\,+ \, 3\, (1,\fund,\antifund)\, +\, 3\,(\antifund,1\fund)
\eeqa
The spectrum can be encoded in the quiver diagram \footnote{A quiver is 
just a picture encoding the chiral spectrum of a theory, with nodes 
representing gauge factors and arrows representing bi-fundamental matter, 
see e.g. \cite{dm,dgm,ks,lnv,uquiver}.} in figure \ref{dthree}b.
Even though it is not realistic, it is a good simple example of a local 
model for an intersecting brane world.

\medskip

{\bf Mirror symmetry interpretation}

This kind of D6-brane configuration has been quite studied in the mirror 
symmetry literature for threefolds $\IW$ corresponding to mirrors of 
complex 
cones over del Pezzo surfaces \cite{hiv,hi,cfikv,fhhi}. In fact the set 
of D6-branes we 
have built maps to D3-branes sitting at the singularity $\IM$. Moreover, 
the D6-brane wrapped on the $\IT^3$ able to move off to infinity is mirror 
to a D3-brane in $\IM$ away from the singular point. The continuous 
deformation of 3-cycles in figure \ref{tadpole} is mirror to moving a 
D3-brane off the singularity. 

The explicit threefold above is mirror to the $\IC^3/\IZ_3$ orbifold 
singularity, so the spectrum on the D6-branes wrapped on the three 
$\IS^3$'s matches that of a set of D3-branes (in $N$ copies of the regular 
representation of $\IZ_3$) at the singularity \cite{dm,dgm,ks,lnv,hu}. It 
is amusing to note that the superpotential of the configuration, which 
arises via open string world-sheet instantons in the D6-brane 
configuration, should match the familiar result from D3-branes at 
singularities \cite{dgm,ks,lnv}.

It is interesting to note that, at the point in complex structure moduli 
space where the D6-brane configuration is supersymmetric, the geometry 
$\IW$ has  a $\IZ_3$ symmetry. It corresponds to the symmetry 
\beqa
z & \to & e^{2\pi i/3} z \nonumber \\
\pmatrix{p \cr q} & \to & \pmatrix{-1 & -1 \cr 1 & 0} \pmatrix{p \cr q}
\label{zthree}
\eeqa
namely a rotation on the $z$-plane times an order three $SL(2,Z)$ action 
on the elliptic fiber (this is simpy an order three geometric action 
$x_1+ix_2 \to e^{2\pi i/3} (x_1+ix_2)$ on the complex coordinate of 
the elliptic fiber). This is the mirror of the point in K\"ahler moduli 
space of $\IM$ that corresponds to the CFT orbifold, and the symmetry is 
the so-called $\IZ_3$ quantum symmetry of the orbifold CFT, not visible 
in classical geometry but present in the $\alpha'$-exact theory
\footnote{There is also a less interesting symmetry, the symmetry between 
the three complex planes in $\IC^3/\IZ_3$ (which also relates the three 
bi-fundamental chiral multiplets on brane probes). A discrete subgroup of 
it seems to be also geometrically realized in the manifold $\IW$ as a 
shift 
of the elliptic fiber (which exchanges the different intersection among 
3-cycles).}.

It is amusing to compare and match the different ingredients in both 
mirror constructions (for instance, the RR tadpoles cancellation 
conditions, the cancellation of mixed $U(1)$ - non-abelian anomalies via 
the Green-Schwarz mechanism, \cite{iru} vs. \cite{afiru}, etc), which had 
been independently derived for D-branes at singularities and D-branes 
wrapped on intersecting cycles. Thus seemingly different mechanisms end up 
related by mirror symmetry. Leaving the discussion as an exercise for the 
interested reader, we turn to construct further models.

Finally we mention that quiver diagrams providing the spectra and 
superpotentials for other similar configurations of D6-branes on compact 
3-cycles in other threefolds $\IW$ (mirror to other cones over del Pezzo 
surfaces) may be found in e.g. \cite{hi}. Instead of reviewing this known 
material, we turn to the discussion of new models, including D6-branes on 
non-compact 3-cycles.

\subsubsection{A Standard Model -like example}
\label{standard}

{\bf Some non-compact 3-cycles}

In this section we would like to consider the same threefold as above, 
but include D6-branes on non-compact 3-cycles. They will correspond to 
global symmetries from the viewpoint of the four-dimensional gauge field 
theory.

\begin{figure}
\begin{center}
\centering
\epsfysize=4.5cm
\leavevmode
\epsfbox{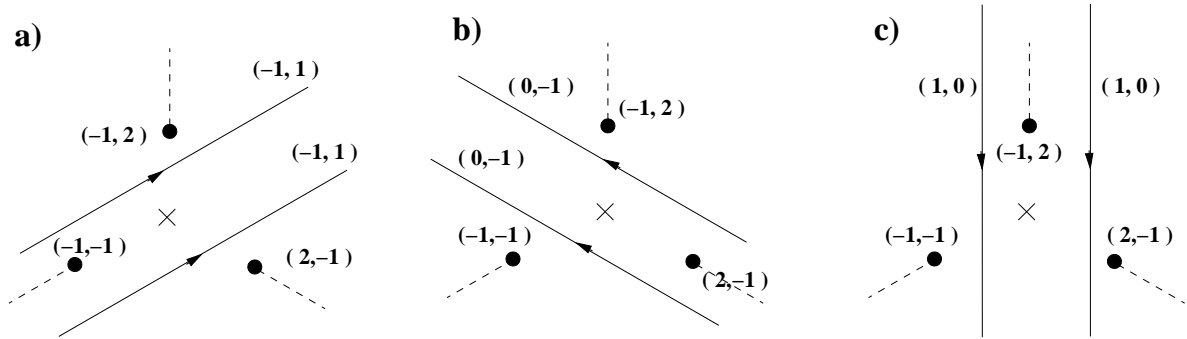}
\end{center}
\caption[]{\small Several useful examples of non-compact 3-cycles.}
\label{noncompact}
\end{figure}

Clearly there is a lot of freedom in choosing non-compact 3-cycles. To 
somehow constrain it, we choose to consider 3-cycles that preserve the 
same $\NN=1$ supersymmetry as the above D6-brane configuration (at the 
$\IZ_3$ symmetric point in complex structure moduli space). We consider 
for instance the non-compact 3-cycles obtained by fibering the non-compact 
direction in $\IC^*$ and the corresponding $(p,q)$ 1-cycles in the 
elliptic fiber over the lines shown in figure \ref{noncompact}. Using the 
conditions in section \ref{supersymmetry}, one may check that these 
3-cycles are supersymmetric, if one chooses the slopes in the $z$-planes 
as in figure \ref{noncompact}.

Let us denote $[\Lambda_{(p,q)}^{\pm}]$ the homology classes of the 
non-compact $(p,q)$ 3-cycles to the right (resp. left), as one moves 
counterclokwise, of the compact 3-cycle which they do not intersect. 
Hence, the classes of the 3-cycles in figure \ref{noncompact}a,b,c are 
$[\Lambda_{(-1,1)}^{\pm}]$, $[\Lambda_{(0,-1)}^{\pm}]$, 
$[\Lambda_{(1,0)}]^{\pm}$. 

The intersection numbers of these non-compact 3-cycles with the basic 
compact 3-cycles are
\beqa
& [\Lambda_{(-1,1)}^+]\cdot [\Sigma_{(2,-1)}]\, =\, 1 \;\; & ;  \;\;
[\Lambda_{(-1,1)}^+]\cdot [\Sigma_{(-1,-1)}]\, =\, [\Lambda_{(-1,1)}^+]
\cdot [\Sigma_{(-1,2)}]\, =\, 0 \nonumber \\
& [ \Lambda_{(-1,1)}^{-} ]\cdot [ \Sigma_{(-1,2)} ]\, =\, -1 \;\; & ; 
\;\;
[\Lambda_{(-1,1)}^{-}]\cdot [\Sigma_{(-1,-1)}]\, =\, 
[\Lambda_{(-1,1)}^{-}]\cdot [\Sigma_{(2,-1)}]\, =\, 0 \;\;
\eeqa
while the remaining can be determined from the $\IZ_3$ symmetry 
(\ref{zthree}).

\begin{figure}
\begin{center}
\centering
\epsfysize=3.5cm
\leavevmode
\epsfbox{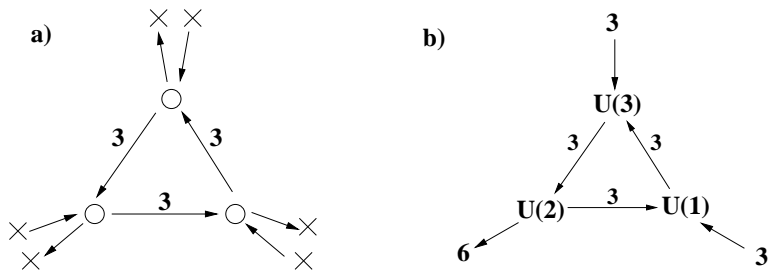}
\end{center}
\caption[]{\small Figure a) shows the quiver diagram for the generic gauge 
theory sector on a set of D6-branes wrapped on the basic compact $(p,q)$ 
3-cycles and on the non-compact 3-cycles in figure \ref{noncompact}. 
Figure b) shows the quiver diagram encoding a chiral gauge theory 
constructed using a particular (tadpole free) choice of wrapped D6-branes; 
the spectrum is remarkably close to that of the Standard Model.}
\label{quiversm}
\end{figure}

Now we would like to consider a general configuration of D6-branes wrapped 
on the compact 3-cycles $[\Sigma_{(p,q)}]$ and on these non-compact 
3-cycles. The spectrum of the resulting four-dimensional gauge field 
theory can be encoded in the quiver diagram fig \ref{quiversm}a), where 
outer nodes correspond to global symmetry groups associated to 
non-compact D6-branes.

\medskip

{\bf Interesection numbers and RR tadpoles}

In order to build consistent configurations these wrapped D6-branes must 
be combined consistently with cancellation of localized RR tadpoles. As 
discussed above, this amounts to requiring the numbers of fundamentals and 
antifundamentals to be equal for any internal node in the quiver in figure 
\ref{quiversm}a. Equivalently, we may take the total homology class 
obtained by wrapping $N_{(p,q)}$, $M_{(p,q)}^{\pm}$ D6-branes on the 
3-cycles $[\Sigma_{(p,q)}]$, $[\Lambda_{(p,q)}^{\pm}]$, and impose the 
condition (\ref{localrr}) to obtain
\beqa
3\, N_{(-1,-1)}\, -\, 3\, N_{(2,-1)}\, + \, M_{(0,-1)}^+ \, -\, 
M_{(-1,1)}^-  \, = \, 0 \nonumber \\
3\, N_{(2,-1)}\, -\, 3\, N_{(-1,2)}\, + \, M_{(1,0)}^+ \, -\, 
M_{(0,-1)}^-  \, = \, 0 \nonumber \\
3\, N_{(-1,2)}\, -\, 3\, N_{(-1,-1)}\, + \, M_{(-1,1)}^+ \, -\, 
M_{(1,0)}^-  \, = \, 0 
\label{rrtad}
\eeqa

For completeness, we would like to present a more geometrical viewpoint on 
the contribution of non-compact 3-cycles to localized RR tadpoles.
Using the intersection numbers with the compact 3-cycles, it is easy to 
obtain certain relations between homology classes, which hold in compactly 
supported homology (that is, up to homology elements non-intersecting the 
compact 3-cycles). For instance
\beqa
&[\Sigma_{(-1,-1)}]\, +\, [\Sigma_{(2,-1)}]\, +\,[\Sigma_{(-1,2)}]\, =\, 0
& \\
&3\, [\Lambda_{(-1,1)}^+] \, +\, 3\, [\Lambda_{(0,-1)}^-] \, =\,
[\Sigma_{(-1,2)}] \quad ; \quad 
3\, [\Lambda_{(1,0)}^+] \, +\, 3\, [\Lambda_{(1,0)}^-] \, =\,
-[\Sigma_{(-1,2)}]  & \nonumber \\
& 3\, [\Lambda_{(0,-1)}^+] \, +\, 3\, [\Lambda_{(1,0)}^-] \, =\,
[\Sigma_{(-1,-1)}]  \quad ; \quad 
3\, [\Lambda_{(-1,1)}^+] \, +\, 3\, [\Lambda_{(-1,1)}^-] \, =\,
-[\Sigma_{(-1,-1)}]  & \nonumber \\
& 3\, [\Lambda_{(1,0)}^+] \, +\, 3\, [\Lambda_{(-1,1)}^-] \, =\,
[\Sigma_{(2,-1)}]  \quad ; \quad 
3\, [\Lambda_{(-1,1)}^+] \, +\, 3\, [\Lambda_{(-1,1)}^-] \, =\,
-[\Sigma_{(2,-1)}]  & \nonumber
\label{relations}
\eeqa

\begin{figure}
\begin{center}
\centering
\epsfysize=4.5cm
\leavevmode
\epsfbox{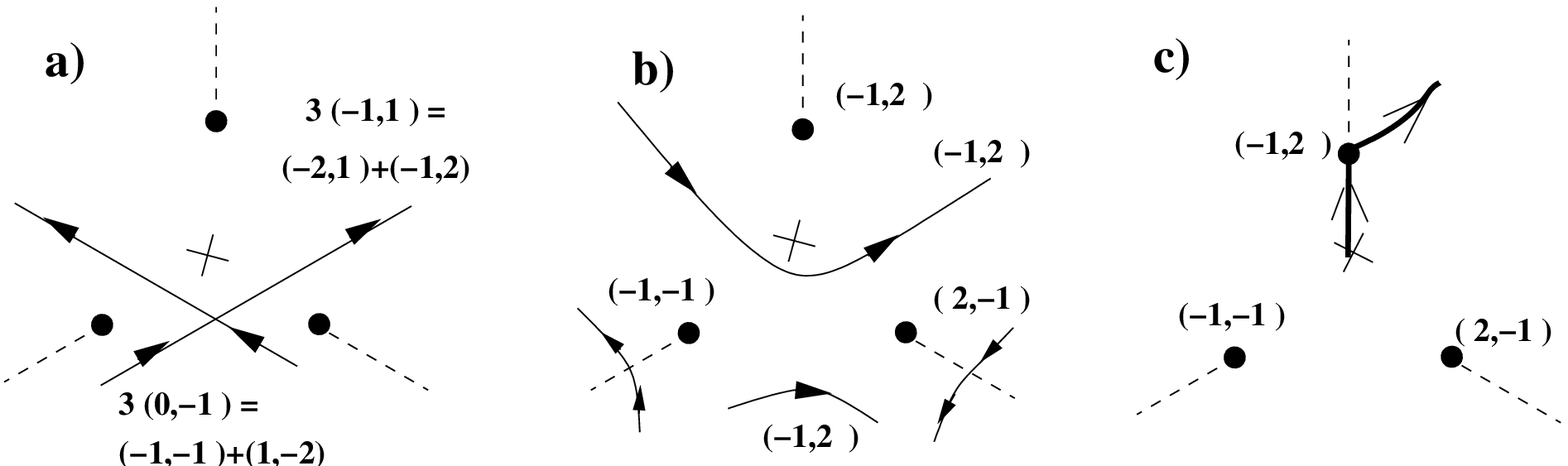}
\end{center}
\caption[]{\small Geometric derivation of the relation 
$3 [\Lambda_{(-1,1)}^+]  + 3 [\Lambda_{(0,-1)}^-]  = [\Sigma_{(-1,2)}]$.
We start in a) from the two sets of non-compact $\Lambda$ 3-cycles; after 
splitting their 1-cycle fiber in the elliptic fibration, recombining the 
$(-1,2)$ segments, and moving $(p,q)$ segments across $(p,q)$ 
degenerations, without prong creation, we reach b); moving the remaining 
$(-1,2)$ segment across the $\IC^*$ degeneration, with prong creation, 
(we use thickness to signal that the prong wraps the $\IS^1$ in $\IC^*$, 
as opposed to the original cycle) leaves a compact 3-cycle 
$[\Sigma_{(-1,2)}]$ plus non-compact 3-cycles not intersecting the 
compact ones.}
\label{tadpole1}
\end{figure}

\begin{figure}
\begin{center}
\centering
\epsfysize=4.5cm
\leavevmode
\epsfbox{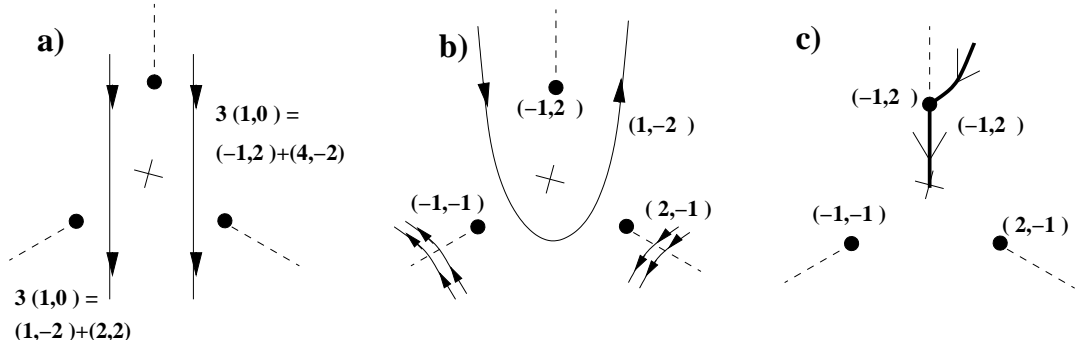}
\end{center}
\caption[]{\small 
Geometric derivation of the relation 
$3 [\Lambda_{(1,0)}^+] + 3\, [\Lambda_{(1,0)}^-] = -[\Sigma_{(-1,2)}]$.
We start in a) from the two sets of non-compact $\Lambda$ 3-cycles; after 
splitting their 1-cycle fiber in the elliptic fibration, recombining the 
$(1,-2)$ segments, and moving $(p,q)$ segments across $(p,q)$ 
degenerations, without prong creation, we reach b); moving the remaining 
$(1,-2)$ segment across the $\IC^*$ degeneration, with prong creation, 
leaves a compact 3-cycle $-[\Sigma_{(-1,2)}]$ plus non-compact 3-cycles 
not intersecting the compact ones.}
\label{tadpole2}
\end{figure}

Alternatively, such relations can be derived by a deformation argument, 
shown in figures \ref{regfract}, \ref{tadpole1}, \ref{tadpole2} for the 
relations in the first and second lines.

Use of relations of this kind would allow a re-derivation of 
(\ref{rrtad}). We prefer instead to use them directly in the example 
below.

\medskip

{\bf The SM like example}

We would like to apply the above ingredients in the construction of a 
local intersecting brane world model with semi-realistic spectrum. A 
simple and very interesting example is provided by combining compact and 
non-compact 3-cycles of the above kinds in a RR tadpole-free fashion as 
indicated in figure \ref{quiversm}b). The numbers in the outer nodes 
denote the number of D6-branes wrapped in the corresponding non-compact 
3-cycles (these need not be overlapping, hence the global symmetry is in 
general abelian, with non-abelian enhancement in non-generic situations 
of overlapping branes). 

One easily verifies that the D6-brane numbers satisfy the localized RR 
tadpole cancellation equations (\ref{rrtad}). Equivalently, one may use 
the relations (\ref{relations}) to show that the total localized tadpole 
vanishes
\beqa
& 3\, [\Sigma_{(-1,2)}]\, +\, 2\, [\Sigma_{(-1,-1)}]\,
+\, [\Sigma_{(2,-1)}]\, +\, 6\, [\Lambda_{(1,0)}^+]\, +\,
3\, [\Lambda_{(-1,1)}^-]\, +\, 3\, [\Lambda_{(1,0)}^-]\, = &
\nonumber \\
&=\, 2\, [\Sigma_{(-1,2)}]\, +\, 2\, [\Sigma_{(-1,-1)}]\, +\,
2\, [\Sigma_{(2,-1)}] \, =\, 0 
\eeqa

The initial four-dimensional gauge group is $U(3)\times U(2)\times U(1)$, 
but is is easy to check that two linear combinations of the $U(1)$'s are 
anomalous (with anomaly cancelled by the Green-Schwarz mechanism) and get 
massive due to the $B\wedge F$ coupling
\footnote{Following \cite{afiru}, these are given in our case by $\sum_a 
N_{(p_a,q_a)} B_a F_{a}$, with $B_a=\int_{\Sigma_{(p_a,q_a)}} C_5$.}. A 
third linear combination
\beqa
Q_Y\, =\, -(\, \frac 13\, Q_3 \, +\, \frac 12 Q_2 \, +\, Q_1\, )
\eeqa
(with $Q_i$ the $U(1)$ generator of the corresponding $U(N)$ factor) does 
not couple to any four-dimensional B-field and hence remains massless. The 
generator $Q_Y$ plays the role of hypercharge in the model, as can be 
checked from the full chiral multiplet spectrum 
\beqa
& SU(3)\times SU(2)\times U(1)_Y & \nonumber \\
& 3(3,2)_{1/6} \, + \, 3({\ov 3},1)_{-2/3} \, +\, 3({\ov 3},1)_{1/3}\, +&
\nonumber \\
&+ \, 3(1,2)_{1/2}\, +\, 3(1,1)_{1}\, +\, 6(1,2)_{-1/2} &
\eeqa
which correctly reproduces left-handed quarks, right-handed quarks, 
leptons and three pairs of vector-like higgs multiplets.

Notice that we have ignored states arising from intersections among 
non-compact 3-cycles, mainly because they are uncharged under the 
four-dimensional gauge group in the non-compact setup. Upon 
compactification, the original non-compact cycles also provide 
four-dimensional gauge interacions; however compactification will in 
general also influence the number of intersection among originally 
non-compact 3-cycles, so it is not too meaningful to study these states in 
a compactification-independent setup.

Clearly, many phenomenological features of the above model could be 
mentioned here. However we prefer to move on and simply mention that the 
main features are similar to those already mentioned in \cite{afiru} 
\footnote{An important particular feature of this model is that it 
presents gauge coupling unification at the string scale {\em if} we sit at 
the $\IZ_3$ symmetric point. Hence unification may be considered 
accidental (not generic over moduli space), or barely natural (since there 
is an enhanced symmetry at such point in moduli space. We would like to 
note in passing that by the mirror symmetry relation below, the same 
statements can be made concerning the models in \cite{aiqu}.}. However, 
notice that, as mentioned in the introduction, the present model avoids 
the problems in lowering the string scale present in the toroidal setup.

It would be interesting to study the possibilities of model building in 
models with non-trivial NS-NS and RR flux. As discussed in \cite{uflux}, 
suitable configurations of fluxes may contribute to RR tadpoles, hence 
replacing some of the D-branes. This may lead to an interesting mechanism 
to potentially getting rid of unwanted gauge factors and/or matter 
multiplets. We leave this interesting possibility for future work.

\medskip

{\bf Mirror symmetry, and further models}

\begin{figure}
\begin{center}
\centering
\epsfysize=3.5cm
\leavevmode
\epsfbox{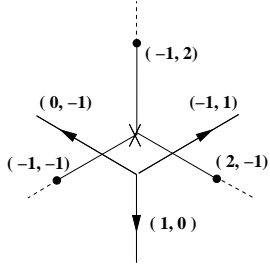}
\end{center}
\caption[]{\small One of the 3-cycles, $\Pi_{(1,0)}$, on which wrapped 
D6-branes reproduce the mirror of D7-branes at the $\IC^3/\IZ_3$ 
singularity. The remaining two are obtained by applying the $\IZ_3$ 
geometric symmetry.} 
\label{dseven}
\end{figure}

Given that the threefold we have used is mirror to the $\IC^3/\IZ_3$ 
orbifold singularity, one may wonder whether the above Standard Model 
D-brane configuration has a construction in terms of D-branes at the 
$\IZ_3$ singularity. Phenomenological model-building with such D-branes 
was studied in \cite{aiqu} (see also \cite{aa}), and it is interesting to 
notice that the above model was not found. The key point is that the 
non-compact D-branes used in \cite{aiqu}, namely D7-branes, even though 
preserved $\NN=1$ supersymmetry, do not coincide with the mirrors of our 
non-compact D6-branes. 

It is a simple exercise to construct D6-branes on non-compact 3-cycles 
which are mirror to D7-branes on the $\IZ_3$ singularity. One such 
3-cycle, denoted $\Pi_{(1,0)}$, is shown in figure \ref{dseven}, while the 
remaining two (denoted $\Pi_{(-1,1)}$, $\Pi_{(0,-1)}$) are obtained 
through action of the $\IZ_3$ geometric symmetry \footnote{One may 
wonder which holomorphic 4-cycle the D7-brane is wrapping in 
$\IC^3/\IZ_3$. To some extent this information is not too explicit in the 
mirror geometry $\IW$, since for example the symmetry between the three 
complex planes (and hence the difference among D7-branes wrapped on 
different 4-cycles $z_i=0$, $i=1,2,3$) seems to be realized simply by a 
shift in the elliptic fiber direction.}. 

\begin{figure}
\begin{center}
\centering
\epsfysize=7.5cm
\leavevmode
\epsfbox{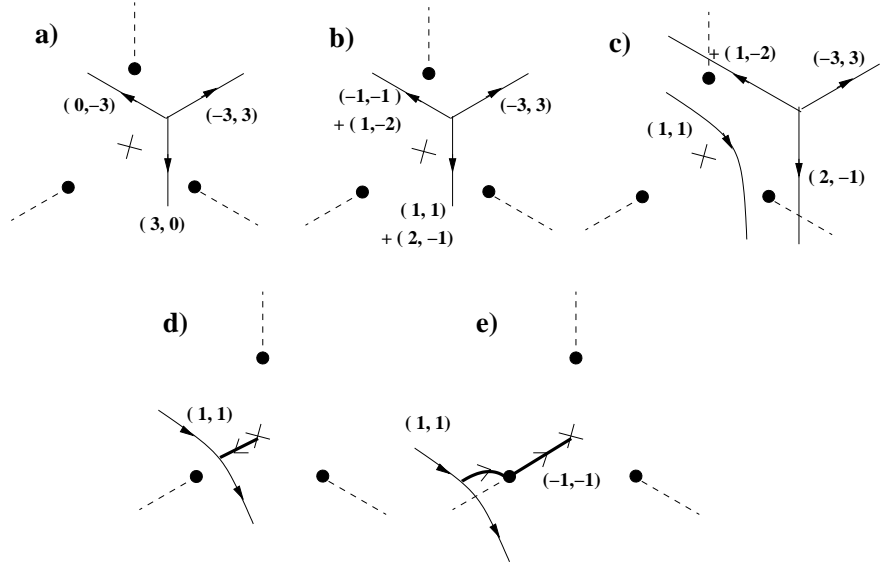}
\end{center}
\caption[]{\small Geometric computation of the contribution of the 
non-compact 3-cycle $\Pi_{(-1,1)}$ to the compact homological charge, 
using by now familiar rules. The deformation argument leads to the 
relation $3 [\Pi_{(-1,1)}] = [\Sigma_{(-1,-1)}]$.}
\label{tadpole}
\end{figure}    

\begin{figure}
\begin{center}
\centering
\epsfysize=3.5cm
\leavevmode
\epsfbox{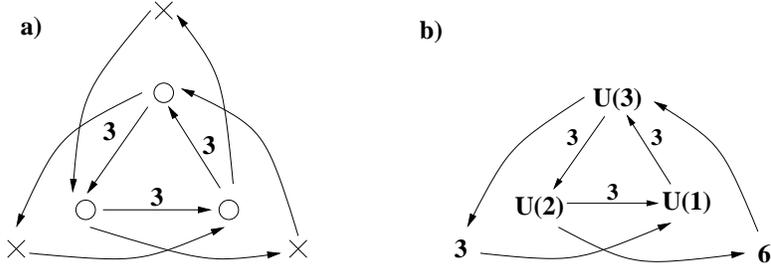}
\end{center}
\caption[]{\small Figure a) shows the quiver diagram for the generic gauge 
theory sector on a set of D6-branes wrapped on 3-cycles providing the 
mirror of D3- and D7-branes at a $\IC^3/\IZ_3$ singularity. Figure b) 
shows the quiver diagram of a Standard Model like D6-brane configuration, 
mirror to a D3/D7 -brane configuration in \cite{aiqu}.}
\label{otherqsm}
\end{figure}

The contribution of such 3-cycles to localized RR tadpoles are easily 
computed by checking intersection numbers with the basic compact 
3-cycles, or by a deformation argument as shown in figure \ref{tadpole}.
The corresponding homology relation may be written as
\beqa
3\, [\Pi_{(-1,1)}] \, =\, [\Sigma_{(-1,-1)}]
\eeqa
Computing intersection number one obtains the generic structure of the 
resulting quiver diagram, which is shown in figure \ref{otherqsm}a. Using 
D6-branes wrapped on the basic compact 3-cycles and these non-compact 
3-cycles it is straightforward to reproduce the models in Sections 3.3, 
3.4 in \cite{aiqu}. One such example, with SM like spectrum is provided in 
figure \ref{otherqsm}b.

We would like however to emphasize that the original model, directly 
constructed in terms of 3-cycles in the threefold $\IW$ (and without 
familiar mirror interpretation) is much simpler and phenomenologically 
more appealing (in particular, it avoids extra vector-like coloured 
triplets). This, in our view, shows that direct model building using local 
configurations of D6-branes on 3-cycles in non-compact threefolds 
(regardless of their mirror interpretation) is an interesting setup to 
explore new phenomenologically realistic D-brane constructions.

\medskip

Nevertheless, we would like to conclude this section by remarking a fact: 
The realization of the same gauge sector within two completely different 
setups, namely D-branes at singularities and intersecting D-branes, which 
are nevertheless related by a string duality (mirror symmetry). It is 
amusing to understand how both constructions manage to reproduce the same 
phenomenological properties (e.g. number of families arising from the 
structure of the singularity on $\IM$ or from the intersection number of 
3-cycles on $\IW$; etc). It is also interesting to understand they lead to 
the same phenomenological implications (e.g. gauge coupling unification at 
the point in moduli space with enhanced global $\IZ_3$ symmetry; 
superpotential couplings; etc). This fact has far reaching philosophical 
implications concerning our view on how string theory might reproduce the 
observed physics. Indeed, as emphasized in \cite{uflux} the right question 
to ask is {\em not} which setup is the one realized in Nature (since all 
of them are related by string dualities, thus symmetries of Nature), but 
which setup provides the best description at the point of moduli space 
where moduli are eventually stabilized. For instance, the intersecting 
D-brane picture is most useful if moduli stabilize in a regime 
where 3-cycles have large sizes, while the picture of D-branes at the 
$\IZ_3$ orbifold singularity may be most useful in the region near the CFT 
orbifold point (where the 3-cycles have stringy size and the intersecting 
brane picture is less useful).

It is easy to construct other examples of local configurations of wrapped
D6-branes leading to interesting gauge field theories. Leaving a more 
detailed exploration for further work, we turn to discussing possible 
generalization of the above threefolds.

\section{Generalizations}
\label{generalizations}

There are several simple generalizations of the above threefolds 
(\ref{wthreefold}). We discuss them briefly in this section, leaving their 
more detailed exploration (which seems very promising) for future 
research.

\subsection{Multiple $\IC^*$ degenerations}
\label{multiple}

A simple generalization of the above ideas is to consider threefolds 
again given by a $\IT^2\times \IC^*$ fibration over a complex plane, with 
$(p,q)$ degenerations of the elliptic fiber as above, but now with several 
points of degeneration of the $\IC^*$ fibration. Such manifolds are 
described by
\beqa
u\, v\, & = & P(z)\, =\, \prod_{i=1}^N\, (z-z_i) \nonumber \\
y^2 \, & = & x^3 \, + \, f(z)\, x \, + \, g(z)
\label{multiplec}
\eeqa
Such manifolds are still Calabi-Yau \footnote{In fact, can be regarded as 
blown-up versions of a $\IZ_N$ quotient of (\ref{threefold}) by an order
$N$ shift in the $\IS^1$ within the $\IC^*$ fibration, $u\to e^{2\pi i/N} 
u$, $v\to e^{-2\pi i/N} v$, hence leaving the original point $u=v=z=0$ 
fixed.}. The rules to construct compact and non-compact 3-cycles are 
analogous to the above one, by fibering over networks of $(p,q)$ segments, 
which now are allowed to have ends on the diverse degeneration 
points of the $\IC^*$ fibration. Intersection numbers between two such 
compact 3-cycles receive contributions only from intersection 
numbers of segments ending on the same $\IC^*$ degeneration point. 
Intersection numbers of compact 3-cycles with 3-cycles spanning the 
non-compact direction in $\IC^*$ may also receive contributions from 
intersections of segments away from the $\IC^*$ degenerations. We skip 
the listing of detailed rules, since they should be clear from experience 
with previous threefolds.

Examples of such models are relatively similar to the constructions we 
have already considered, hence we turn to other possible generalizations.

\subsection{Double elliptic fibration}
\label{doublef}

Another possible generalization of the threefolds (\ref{wthreefold}) is to 
consider a double elliptic fibration over the $z$ plane
\beqa
y^2 \, & = & x^3 \, + \, f(z)\, x \, + \, g(z) \nonumber \\
{\tilde y}^2 \, & = & {\tilde x}^3 \, + \, {\tilde f}(z)\, {\tilde x} \, + 
\, {\tilde g}(z)
\label{doublefibr}
\eeqa
Namely the threefold is given by two different elliptic fibrations over 
the $z$-plane. At points in the $z$-plane, one of the two elliptic fibers 
$\IT^2$, ${\bf \tilde T}^2$, degenerates due to the pinching of a $(p,q)$ 
or $({\tilde p},{\tilde q})$ 1-cycle. For suitable choices of elliptic 
fibrations the above non-compact threefold is Calabi-Yau. A 
global version of the above double elliptic fibration manifold has 
appeared in \cite{vafa}.

The basic compact 3-cycles in the manifold (\ref{doublefibr}) are obtained 
by considering a segment stretching from a $(p,q)$ to a $({\tilde p}, 
{\tilde q})$ degeneration points, and fibering over it the corresponding 
$(p,q)$ and $({\tilde p},{\tilde q})$ 1-cycles in the elliptic fibrations.

More generally, it is possible to consider 3-cycles arising from networks 
of segments in the $z$-plane, over which we fiber a two-cycle in the 
$\IT^2\times \IT^2$ fiber. One important subtlety is that the charge that 
should be conserved at junctions is the 2-homology class
\beqa
[\Pi_{\rm 2-cycle}] & = & (p[a]+q[b])\, \otimes \, ({\tilde p} 
[{\tilde a}] + ({\tilde q} [{\tilde b}])\, =\, \nonumber \\
& = & p{\tilde p} \, [a]\otimes [{\tilde a}] \, +\,
p{\tilde q} \, [a]\otimes [{\tilde b}] \, +\, 
q{\tilde p} \, [b]\otimes [{\tilde a}] \, +\, 
q{\tilde q} \, [b]\otimes [{\tilde b}] \, +\, 
\eeqa
Which in general is not addition of $(p,q)$ and $({\tilde p}, {\tilde q})$
labels independently. Finally, the external legs in the network should 
have factorized 2-cycles fibered over them, and should end on the 
corresponding $(p,q)$ or $({\tilde p},{\tilde q})$ degenerations.

The intersection numbers between two 3-cycles are computed by careful 
addition with sign of the intersection numbers among the different 
segments. There are two contributions to the intersection numbers of 
segments: i) two segments with labels $(p,q)$, $({\tilde p},{\tilde q})$ 
and $(p',q')$, $({\tilde p},{\tilde q})$ ending on the same $({\tilde 
p},{\tilde q})$ degeneration contribute $I=pq'-qp'$ to the intersection 
number (and similarly for exchanged roles of tilded and untilded); ii) an 
intersection in the $z$-plane of two segments with (in 
general non-factorizable) 2-cycles $[\Pi]$, $[\Pi']$, in the 
$\IT^2\times\IT^2$ fibered over them, contributes $I=[\Pi]\cdot [\Pi']$ to 
the intersection number (with $\cdot$ denoting intersection product in the 
$\IT^2\times\IT^2$ fiber).

Recalling that a $\IC^*$ fibration near a degeneration point provides a 
local model for an elliptic fibration near a degeneration point, we notice 
that the geometries in section \ref{multiple} provide a local model for 
double elliptic fibrations where all degenerations of one fibration are 
mutually local, namely of the same $({\tilde p},{\tilde q})$ type. In fact 
it is easy to recover the 3-cycles and their rules in section 
\ref{multiple} by particularizing the ones in this section.

It is also a simple matter to generalize the discussion about 
supersymmetry in section \ref{supersymmetry} to the present context. For 
instance, a calibrating 3-form can be taken as
\beqa
[\Omega_3 ], =\, (d{\tilde x}_1+id{\tilde x}_2) \wedge (dx_1+idx_2)\wedge 
dz
\eeqa
The basic 3-cycles we study are made of pieces of the form
\beqa
x_1+ix_2 & = & p+\tau q\, =\,a e^{i\theta_1} \nonumber \\
{\tilde x}_1 + i {\tilde x}_2 & = & {\tilde p} + {\tilde \tau} {\tilde q} 
\, =\, b e^{i\theta_2} \nonumber \\
z & = & c e^{i\theta_3}
\eeqa
parametrized by $a,b,c\in \IR$, so that the cycle is slag if each segment 
satisfies $\theta_1+\theta_2+\theta_3=0$.

\begin{figure}
\begin{center}
\centering
\epsfysize=6cm
\leavevmode
\epsfbox{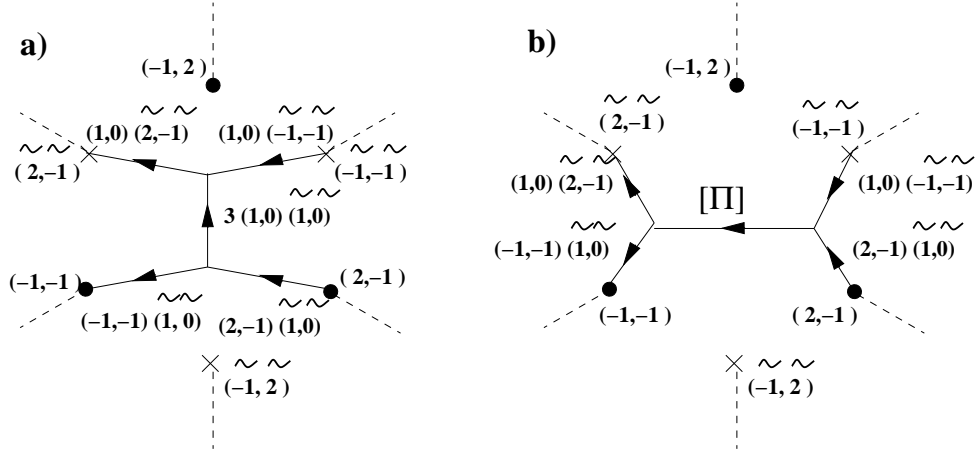}
\end{center}
\caption[]{\small Figure a) shows a 3-cycle in a double elliptic fibration 
threefold. Note the conservation of 2-homology charge in the $\IT^2\times 
\IT^2$ fiber as one moves across junctions. Figure b) shows another 
representative 3-cycle in the same homology class, which is slag in the 
$\IZ_3$ symmetric configuration. Notice that one of the segments has a 
non-factorizable 2-cycle (in the class $[\Pi]=[a_1][a_2]- [a_1][b_2] 
-[a_2][b_1]$) in the $\IT^2\times \IT^2$ fiber.}
\label{double}
\end{figure}

Notice that at some junctions, factorizable 2-cycles in the double 
elliptic fiber may combine into a non-factorizable 2-cycle. We expect that 
the latter will lead to a slag segment if the former do. A particular 
example of this is given below (figure \ref{double}b).

\medskip

We would like to conclude by giving a simple example of a D6-brane 
configuration in such a double elliptic fibration. Consider each double 
elliptic fibration to have the following set of degenerations: $(2,-1), 
(-1,2), (-1,-1)$, which we order in a counterclockwise fashion as in 
figure \ref{double} (this implies a choice of point in the moduli space). 
This space is a Calabi-Yau threefold.
Notice that at this point in moduli space there is a geometric $\IZ_3$ 
symmetry similar to (\ref{zthree}), with the monodromy matrix now acting 
on both elliptic fibers.

In Figure \ref{double}a) we present one particular compact 3-cycle in this 
threefold. Notice that the 2-homology charge in the $\IT^2\times \IT^2$ 
fiber is conserved across junctions. At the $\IZ_3$ symmetric 
configuration, the 3-cycle in figure \ref{double}a) has a slag 
representative shown in figure \ref{double}b). Two other compact 3-cyles 
may be obtained by acting with the $\IZ_3$ geometrical symmetry. 

Let us consider a set of $N$ D6-branes wrapped on each of these 
three compact 3-cycles. It is possible to check that the resulting 
configuration cancels all localized RR tadpoles \footnote{A simple proof
is to start with a homologically trivial small loop at the center of the 
configuration, deform it to grow prongs to and from the elliptic 
degenerations, shrink the initial small loop, and then suitably recombine 
the prongs to reproduce the three 3-cycles. By similar arguments, the 
general RR tadpole condition for this threefold can be shown to be that 
the net number of outgoing prong from the $i^{th}$ elliptic degeneration 
should be $i$-independent.}. Finally by computing 
intersecting numbers, the final four-dimensional chiral gauge theory has 
the quiver in figure \ref{dthree}b. Hence our present example gives a 
different construction of the chiral gauge field theory in section 
\ref{trinification}. It is then clear that double elliptic 
fibrations have enough richness to yield interesting local configurations 
of wrapped D6-branes.

\subsection{Higher genus fibrations}

There is a final class of generalization that we would like to mention. 
These new geometries are obtained by fibering over a complex plane a 
$\IC^*$ fibration and a fibration of a higher genus curve $\Sigma_g$. Over 
certain points in the base $z$-plane a 1-cycle of the genus $g$ curve 
fiber degenerates,  hence degeneration points are labeled by the 
1-homology class of the degenerating 1-cycle $[\Pi]=\sum_{r=1}^g (p_r 
[a_r] + q_r[b_r])$.

In analogy with previous examples, one can construct a set of compact 
3-cycles by taking segments in the $z$-plane stretching between the 
degeneration point of the $\IC^*$ fibration and one of the degeneration 
points of the $\Sigma_g$ fibration, and fibering over it the $\IS^1$ in 
the $\IC^*$ fiber and the corresponding 1-cycle $[\Pi]$ in $\Sigma_g$. 
More generally, one can construct more complicated 3-cycles by 
considering networks of segments in the $z$-plane, over which one 
fibers the $\IS^1$ in the $\IC^*$ fiber and a 1-cycle $[\Pi]$ in 
$\Sigma_g$, and ensuring conservation of the 1-homology class in 
$\Sigma_g$ across junctions in the network.

The intersection number between two such 3-cycles is given by adding 
contributions from segment intersection numbers. The latter only receive 
contributions from two segments, of $[\Pi]$, $[\Pi']$ kinds,
ending on the $\IC^*$ degeneration point. The contribution is given by 
\beqa
[\Pi]\cdot [\Pi']\, =\, (\vec{p};\vec{q})\cdot {\bf I} \cdot 
(\vec{p}\, ';\vec{q}\, ')\, =\, \vec{p}\cdot \vec{q}\, ' - 
\vec{q}\cdot\vec{p}\, '
\eeqa
where $(\vec{p};\vec{q})=(p_r;q_r)$ and ${\bf I}=\pmatrix{0 & \id_g \cr 
-\id_g & 0}$ is the intersection form in $\Sigma_g$.

\medskip

\begin{figure}
\begin{center}
\centering
\epsfysize=4cm
\leavevmode
\epsfbox{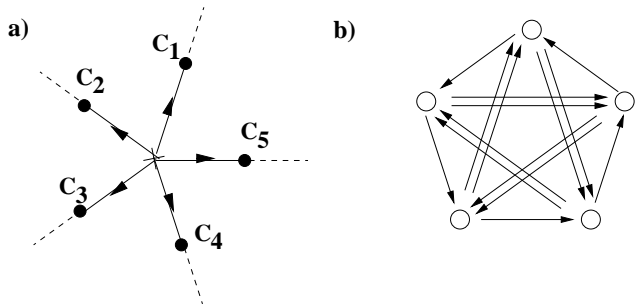}
\end{center}
\caption[]{\small Figure a) shows a configuration of D6-branes wrapped on 
the basic compact 3-cycles in a Calabi-Yau threefold given by a 
$\IC^*$-fibration and a genus 2 curve fibration over the complex plane. 
Figure b) gives the quiver associated to the corresponding chiral gauge 
field theory.}
\label{genus}
\end{figure}

A few particular examples of Calabi-Yau manifolds of this kind have been 
considered in the literature \cite{ray, fhhi}. Their introduction is 
motivated from consideration of the manifolds mirror to orbifold 
singularities $\IC^3/\IZ_N$, other than $\IZ_3$. Indeed, the particular 
case of the mirror of the $\IZ_5$ geometry was obtained from the mirror 
map in \cite{ray}, and is given by a genus 2 curve fibration with a 
set of five degenerate fibers, with labels
\beqa
& C_1=(0,0;-1,2) \quad ; \quad C_2=(-1,0;0,-3) \quad ; \quad 
C_3=(0,1;4,-1) & \nonumber \\ 
& C_4=(0,-1;-1,0) \quad ; \quad C_5=(1,0;-2,2) &
\eeqa
The resulting set of basic compact 3-cycles and the quiver for the chiral 
gauge theory obtained upon wrapping D6-branes on them are shown in figure 
\ref{genus}. Notice that it reproduces the gauge theory on a regular 
D3-brane at the $\IC^3/\IZ_5$ singularity, a system related to ours by 
mirror symmetry.

Certain features of other examples, as well as a discussion of the 
different compact 3-cycles in such geometries, have appeared in 
\cite{fhhi}.

Even though the explicit example above was provided by mirror symmetry, it 
is clear that given a threefold of the kind described above one may use 
D6-branes wrapped on different sets of 3-cycles to generate other 
chiral gauge sectors. However, the constructions of such manifolds and 
cycles is relatively technical and lies beyond the scope of the present 
paper. Clearly, further generalizations by introducing further 
degenerations of the $\IC^*$ fibration, or extending it to a new elliptic 
fibration should be clearly tractable from the constructions above.

\subsection{Orientifold models}

It is straightforward to inmplement orientifold projections in the 
threefolds in section \ref{sthreefold} or their generalizations, whenever 
they have a geometric $\IZ_2$ symmetry. We will not explore the different 
possibilities of model building in detail, but we expect them to be very 
rich, and very likely to yield the Standard Model intersection numbers of 
\cite{imr} in suitable examples. 

Any threefold with a $\IZ_2$ geometric symmetry $g$ may be orientifolded 
by quotienting the configuration by $\Omega g$ (times $(-1)^{F_L}$ if 
required). The configuration of wrapped D6-branes should be $\IZ_2$ 
invariant, via the introduction of orientifold images, and should satisfy 
RR tadpole cancelation conditions that now may include the contribution 
from the orientifold plane charges.

We would like to point out that, if one is not interested in supersymmetric 
configurations, the non-compact setup allows in principle to construct 
models with orientifold planes which are not O6-planes, since the 
corresponding RR charges can escape to infinity. This possibility was 
not available in the compact setup, so the present framework may allow for 
more freedom \footnote{On the other hand, one should note that if the 
O-planes introduced span too many non-compact directions they may lead to 
other localized RR tadpoles requiring its own set of D-branes. However 
the latter typically would not lead to chiral matter on the D6-brane 
gauge theory, so the net effect is similar to simply forgetting their 
tadpoles.}.

There exist examples of such $\IZ_2$ symmetries in particular examples of 
threefolds. To give an example, if all the elliptic fibration degenerations
are mutually local (say all of $(1,0)$ type), and are distributed $\IZ_2$ 
symmetrically with respect to the origin in the $z$-plane, we may mod out 
by $\Omega$ times
\beqa
z\to -z \quad ; \quad x_1\to x_1 \quad ; \quad x_2 \to -x_2 \quad ; \quad 
u\to -u \quad ; \quad v\to -v
\eeqa
The action on $x_1+ix_2$, the complex coordinate on the elliptic fiber, is 
equivalent to $y\to {\ov y},\, x\to {\ov x}$ in (\ref{wthreefold}). The 
orientifold planes span 
a complex curve in $u,v,z$ times a real line in $x_1+ix_2$, so the model 
contains O4-planes. Their RR flux may escape to infinity, so there are no 
tadpoles cancellation condicions associated to them.
Other special examples of $\IZ_2$ symmetries in particular threefolds are 
easily constructed and may be useful in model building.

Clearly there are many possibilities. This kind of projections would break 
the supersymmetries of the configuration, but many models in the 
literature are already non-supersymmetric (in the spirit of \cite{nonsusy}). 
Another interesting fact is that most of these orientifold will not 
contribute to the tadpoles for the RR fields arising from integrals of 
the RR 7-form over compact 3-cycles, hence they do not modify the RR 
tadpole conditions for D6-branes. 
Indirectly (via the relation between tadpoles and anomalies) this suggests 
that the matter content on the gauge theory may be free of two-index 
tensor representations of the gauge groups (or may less fortunately lead 
to equal number of symmetric and antisymmetric representations). This is 
important for phenomenological purposes, since such fields tend to yield 
exotics not present in the Standard Model (see e.g. \cite{susy}).

\medskip

There are also the more familiar orientifold projections yielding 
O6-planes wrapped on 3-cycles of the kind studied above. These are 
obtained when the geometric $\IZ_2$ acts antiholomorphically on the 
complex coordinates of the threefold. It is straighforward to find 
examples of such actions.

For instance, starting with (\ref{wthreefold}) with elliptic degenerations 
$(2,-1),(-1,2),(-1,-1)$ in the $\IZ_3$ symmetric configuration, we may mod 
by
\beqa
u\to {\ov v} \quad ;\quad v \to {\ov u} \quad ; \quad z\to {\ov z} \quad ; 
\quad x_1 \to -x_1 \quad ; \quad x_2\to x_2
\eeqa
where $x_1+ix_2$ is the complex coordinates in the $\IT^2$ fiber. The 
O6-plane spans the imaginary axis in the $z$-plane, the $\IS^1$ in $\IC^*$ 
and the 1-cycle $(1,-2)$ in the elliptic fiber. Hence it contributes an 
amount of $-4[\Sigma_{(1,-2)}]$ to the localized RR tadpole, with the $-4$ 
arising from the O6-plane charge.

Let us build a configuration of wrapped D6-branes in this orientifolded 
threefold background. Wrapping $N$ D6-branes on the invariant 3-cycle 
$[\Sigma_{(-1,2)}]$, and $N+4$ D6-branes on the ($\IZ_2$-related) 3-cycles
$[\Sigma_{(2,-1)}]$, $[\Sigma_{(-1,-1)}]$, the configuration is  
tadpole-free. The gauge theory spectrum is
\beqa
{\rm Gauge}\, {\rm  group} \quad & \quad \quad \quad SO(N)\times 
U(N+4) \nonumber \\
\NN=1 \,\, {\rm Ch.} \,\,  {\rm Mult.} \quad & \quad 
3\,(\fund,\antifund,1) \,+ \, 3\, (1,\Yasymm)
\eeqa
which can be codified in an orientifolded quiver \cite{uquiver}.
Choosing $N=1$, we recover a three-family $SU(5)$ grand unified model. In 
fact, this is nothing but the mirror of a (local version of a) model in 
\cite{lpt}.

Hoping that these examples suffice to observe the general rules in 
orientifold constructions, we leave the detailed exploration of other 
explicit examples for future research. See one further example in section 
\ref{twofold}.

\section{Local intersecting brane-worlds in a two-torus times a 
non-compact two-fold}
\label{twofold}

In this section we describe the construction of yet another class of 
non-compact threefolds. They are simpler than the above ones, 
yet we have preferred to postpone their discussion since in a sense
they have a less general structure. 

The manifolds we would like to consider are a global product of a 
two-torus times a non-compact two-fold $\IX$. In principle we would like 
to restrict to the Calabi-Yau situation, which enforces $\IX$ to be 
either flat or a piece of a K3. For convenience, and also because it 
parallels previous analysis, we choose to describe our two-fold as an 
elliptic fibration over a complex plane.
\beqa
y^2 \, & = & x^3 \, + \, f(z)\, x \, + \, g(z)
\label{piecekthree}
\eeqa
Note that the geometry is very similar to the double elliptic fibrations 
in section \ref{doublef}, with one elliptic fibration being trivial, i.e. 
without degeneration points. Hence the rules below are easily obtained 
by particularizing those in section \ref{doublef}.

On this geometry we can consider different kinds of 3-cycles. The 
basic ones can be constructed by taking the product of a 1-cycle $(m,n)$ 
in the two-torus, times a segment in the $z$-plane, over which we fiber a 
1-cycle $(p,q)$ in the elliptic fibration. We will describe this 
pictorially by a segment in the $z$-plane carrying two labels $(m,n)$ and 
$(p,q)$. For the 3-cycle to be closed, we need both ends of the segment to 
end on $(p,q)$ degenerations of the elliptic fibration.

A slight generalization is to consider segments over which we fiber a 
two-cycle in the homology class $[\Pi]$ in $\IT^2\times\IT^2$, where the 
last factor is the elliptic fiber. For non-factorizable $[\Pi]$ such 
3-cycles are non-compact, since they are unable to end on degeneration 
points of the elliptic fibration.

As usual, we may also consider more complicated 3-cycles, obtained by 
considering a network of $[\Pi]$ segments over the $z$-plane, over which 
we fiber the corresponding 2-cycles in $\IT^2\times \IT^2$, and ensuring 
conservation of the 2-homology class $[\Pi]$ across junctions. In order to 
obtain compact 3-cycles, the external legs in the network must have 
factorizable $[\Pi]$ and must end on the corresponding $(p,q)$ 
degeneration of the elliptic fibration.

The intersection number of two 3-cycles receives two kinds of 
contributions from segment intersection numbers: i) two segments with 
labels $(m,n)$, $(p,q)$, and $(m',n')$, $(p,q)$ ending on the same 
$({\tilde p},{\tilde q})$ degeneration contribute $I=mn'-nm'$ to the 
intersection number; ii) an intersection in the $z$-plane of two segments 
with (in general non-factorizable) labels $[\Pi]$, $[\Pi']$, contributes 
$I=[\Pi]\cdot [\Pi']$ to the intersection number (with $\cdot$ denoting 
intersection product in $\IT^2\times\IT^2$).

Notice that factorizable 3-cycles contain at least one non-trivial element 
in its first homology group (since they are a product of $\IS^1$ in 
$\IT^2$ times a 2-cycle in $\IX$). Hence, D6-branes on such 3-cycles lead 
to at least one adjoint chiral multiplet.

\medskip

For simplicity, below we consider some examples of this kind of 
construction for a particularly simple kind of elliptic fibration, where 
all degenerations are of the same $(p,q)$ kind, which without loss of 
generality we take to be $(1,0)$. Other case may be analyzed using the 
above rules.

An interesting feature of the elliptic fibration with $N$ $(1,0)$ 
degenerations is that it provides the mirror manifold of a $\IC^2/\IZ_N$ 
ALE singularity \footnote{In fact such singularities are self-mirror, 
 so it is also true that the fibration $\IX$ is a blown-up version of
$\IC^2/\IZ_N$. We prefer to regard it as its mirror to make it consistent 
with the mapping of branes mentioned below.}. Hence the models with 
D6-branes wrapped on compact 3-cycles may be related via two-fold mirror
symmetry with models of D-branes wrapped on 1-cycles in $\IT^2$ and on 
holomorphic cycles in the $\IC^2/\IZ_N$ singularity. In particular cases, 
such B-type branes can be described as D-branes sitting at the singular 
point in $\IC^2/\IZ_N$. Hence the models of D6-branes on compact 3-cycles 
in the threefold may be in some cases related to models of D4-branes 
wrapped on 1-cycles on $\IT^2$ and sitting at a $\IC^2/\IZ_N$ singularity. 
The latter models were introduced in \cite{afiru} (see 
also \cite{cim,otherbrane}). In this sense, our construction in this 
section provides a generalization of such models, including configurations 
not related to D4-brane models. In particular we would 
expect that the general case of D6-branes on 3-cycles on $\IT^2\times \IX$ 
to avoid certain unwanted features of D4-branes models, like the generic 
presence of tachyons at intersections.

\begin{figure}
\begin{center}
\centering
\epsfysize=3.5cm
\leavevmode
\epsfbox{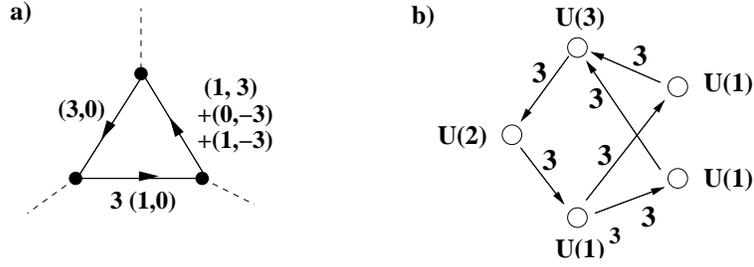}
\end{center}
\caption[]{\small Local configuration of D6-branes wrapped on compact 
3-cycles in a non-compact manifold $\IT^2\times \IX$. The 3-cycles are
products of 1-cycles $(m,n)$ in $\IT^2$ (these are the labels in the 
figure) times 2-cycles in $\IX$. The latter are obtained by fibering over 
each segment the $(1,0)$ 1-cycles in the elliptic fibration, since this is 
the only kind of degeneration. Notice the change of notation with respect 
to other figures, namely the $(m,n)$ labels on the segments refer to the 
1-cycle in $\IT^2$ rather than to on the elliptic fiber.}
\label{twofoldone}
\end{figure}

In figure \ref{twofoldone}a we show a particular choice of 
D6-branes wrapped on compact 3-cycles in $\IT^2$ times an elliptic 
fibration with three $(1,0)$ degenerations. Cancellation of 
localized RR tadpoles is easily checked by a deformation argument. The 
spectrum of the resulting chiral gauge theory is shown in the quiver 
diagram of figure \ref{twofoldone}b. The configuration precisely 
reproduces the D4-brane model in section 4.3 in the first reference in 
\cite{afiru}, to which it is related via mirror symmetry in $\IX$. Notice 
also that certain features like the generic presence of tachyons can be 
tracked to the impossibility of making the different overlapping 3-cycles 
simultaneously slag, no matter the location of the elliptic fibration 
degenerations (i.e. no matter at which point in moduli space one is 
sitting).

Notice that the spectrum is quite close to that of the Standard Model.
An important issue is to obtain a massless hypercharge. As in \cite{afiru} 
one could in principle claim that there is an anomaly free linear 
combination of $U(1)$'s which yields correct hypercharge asignments. 
However, such combination does not pass the test (subsequently brought to 
attention in \cite{imr}, but still too often overlooked in the literature)
of having no $B\wedge F$ couplings with four-dimensional 2-form fields. 
Hence the model has no hypercharge candidate and should be regarded as a 
simple toy model for our construction. This seems to be a general feature 
of models of D6-branes wrapped on only compact 3-cycles in $\IT^2\times 
\IX$. Introduction of non-compact 3-cycles presumably allows to avoid 
such problems in more involved models.

\begin{figure}
\begin{center}
\centering
\epsfysize=3.5cm
\leavevmode
\epsfbox{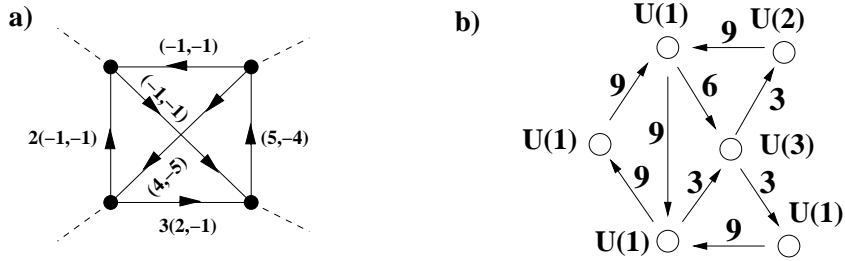}
\end{center}
\caption[]{\small Yet another example of configurations of D6-branes on a 
threefold $\IT^2\times \IX$. As in figure \ref{twofoldone}, labels in 
figure a) refer to the $(m,n)$ charge in $\IT^2$, whereas the $(p,q)$ 
charge in the elliptic fiber is $(1,0)$ for all cycles. Figure b) shows 
the quiver diagram for the resulting gauge theory.}
\label{twofoldtwo}
\end{figure}

In figure \ref{twofoldtwo} we present another example of wrapped D6-brane
configuration, which is not mirror of any configuration of D4-branes.
It yields a gauge theory Standard Model like group and three chiral 
families, but unfortunately also no hypercharge candidate (and plenty of 
extra matter). It should not be difficult to modify examples of this kind 
e.g. by introducing non-compact D6-branes and build phenomenologically more 
appealing models.

\medskip

\begin{figure}
\begin{center}
\centering
\epsfysize=4.5cm
\leavevmode
\epsfbox{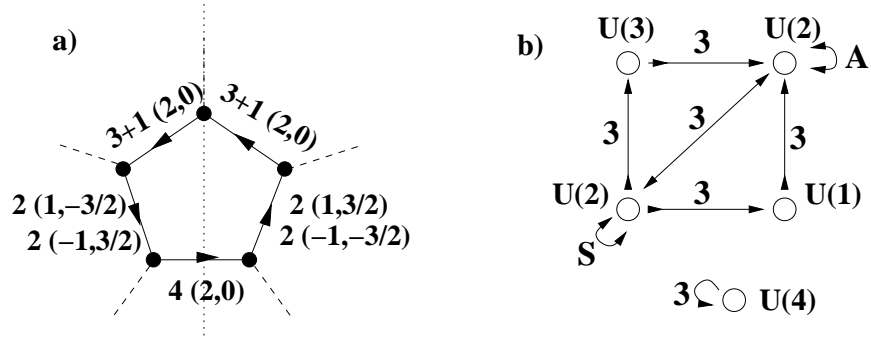}
\end{center}
\caption[]{\small 
A simple example of D6-branes wrapped on 3-cycles in an orientifold of 
the $\IT^2\times \IX$ geometry, along with its quiver diagram. Conventions 
are as in other figures in this section. Note that quiver arrows may have 
two outgoing ends in an orientifold model \cite{uquiver}. Also, {\bf S} 
and {\bf A} denote two index symmetric and antisymmetric representations.}
\label{orientifold}
\end{figure}

Clearly it is possible to implement orientifold projections in the present 
setup. One of the simplest possibilities is to consider only $(1,0)$ 
degenerations of the elliptic fibration, and distribute them in a $\IZ_2$ 
invariant fashion in the $z$-plane (say, $z\to -{\ov z}$). One may then 
mod out by $\Omega$ times
\beqa
z\to -{\ov z} \quad ; \quad x_1 \to x_1 \quad ; \quad x_2\to -x_2 \quad ; 
\quad u\to -{\ov v} \quad ; \quad v\to {\ov u}
\eeqa
This introduces O6-planes wrapped on the $\IS^1$ in $\IC^*$, the imaginary 
axis in the $z$-plane, and the $(1,0)$ 1-cycle in the elliptic fiber. The 
contribution of the O6-plane to the RR tadpole cancellation is computable 
from deformation arguments, or checking intersection numbers. Such 
orientifold geometries are typically mirror of orientifolds of 
$\IT^2\times \IC^2/\IZ_N$ orbifolds. Inclusion of wrapped D6-branes (and 
their $\IZ_2$ images) lead to intersecting brane worlds, which may have a 
simple mirror or not (as models of D4-branes at the orbifold geometry), 
depending on the kind of wrapped branes used. Rules to compute the 
spectrum follow from standard arguments and it would be lengthy to 
attempt a general list here. Let us instead present an illustrative exaple 
of a simple tadpole free configuration is given in figure 
\ref{orientifold}. This particular example contains branes with simple 
mirror D4-brane interpretation; indeed the model is mirror to the 
D4-brane model in appendix I in \cite{cim}. It corresponds to a left-right 
symmetric model $SU(3)\times SU(2)_L \times SU(2)_R\times U(1)_{B-L}$ with 
three families of quarks and leptons.

\section{Configurations with Standard Model intersection numbers}
\label{imr}

\subsection{Construction of the configurations}

In this Section we would like to address a last issue. In \cite{imr}
a set of intersection numbers for 3-cycles in orientifolded Calabi-Yau 
models were proposed, such that, when wrapped by D6-branes, lead to a 
gauge sector with exactly the chiral fermion content of the standard 
model. Particular realizations of 3-cycles with those intersection numbers 
has been achieved in orientifolds of the six-torus \cite{imr}, certain 
orbifolds \cite{cim} and the quintic Calabi-Yau threefold \cite{bbkl}. 

It would be extremely interesting to realize such intersection 
numbers in non-compact threefolds of the kind studied in this paper. A 
superficial direct search has not led to this structure, but we certainly 
do expect such models to exist, and be classifiable once more formal 
exploration tools are developed. The threefolds we have studied seem rich 
enough to allow intersection patters as specific as those in \cite{imr}.
We expect much progress in this direction.

In this Section we introduce a new class of non-compact threefolds, which 
are not Calabi-Yau, but have the interesting property of leading, in a 
relatively simple way to the Standard Model intersection numbers in 
\cite{imr}. The threefolds we consider are topologically $\IT^2\times 
\IT^2\times \IY$, with $\IY$ a non-compact Riemann surface. The motivation 
to use these threefold will become clear below. Since the geometry of the 
onefold $\IY$ and its set of compact 1-cycles are quite different from 
previous geometries, we briefly describe them.

\begin{figure}
\begin{center}
\centering
\epsfysize=3.5cm
\leavevmode
\epsfbox{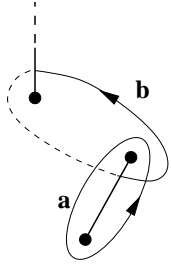}
\end{center}
\caption[]{\small Double cover representation of the two-torus (minus one 
point). Thick lines correspond to branch cuts joining the two sheets. Thin 
lines show the two non-trivial $a$ and $b$ 1-cycles; continuous and 
dashed pieces live in different sheets of the cover.}
\label{twotorus}
\end{figure}

Any non-compact Riemann surface can be conveniently described as a double 
cover of the complex $z$-plane, with a certain number $N$ of branch points 
$z_i$
\beqa
y^2 \, = \, \prod_{i=1}^N \, (z-z_i)
\label{doublecover}
\eeqa
Branch cuts connecting the two sheets of the cover emanate from the branch 
points, and end on other branch cuts or go off to infinity. The double 
cover representation of the 2-torus (minus one point to make it 
non-compact) is shown in figure \ref{twotorus}. 

In this geometry, non-trivial 1-cycles are described by closed paths 
surrounding two branch points. If both share a common branch cut, the 
1-cycle lives in one of the sheets; whereas if the two branch points 
correspond to different branch cuts the resulting 1-cycle crosses them and 
contains pieces in both sheets. Other 1-cycles may be constructed by 
combining the above basic set. Let us note for completeness that the 
1-cycles can be deformed, preserving the homology class, keeping in mind 
to change the sheet whenever they cross a branch cut. Examples of the 
non-trivial $a$ and $b$ 1-cycles in the 2-torus are shown in figure 
\ref{twotorus}. Hence the number of branch points determines the number 
of holes in the Riemann surface, whereas the number of boundary 
components is determined by the structure of branch cuts going off to 
infinity. For simplicity we consider double covers where the number of 
branch cuts going to infinity is minimal, namely zero or one for even or 
odd $N$ (other cases are similar). We also choose a canonical 
representation, to be used in topological matters, where branch points are 
counterclockwise ordered in a circle in the $z$-plane, with branch cuts 
starting at $z_{2k-1}$ and ending at $z_{2k}$ (and a last branch cut 
going off to infinity outwards from the unpaired branch point for odd 
$N$).

\begin{figure}
\begin{center}
\centering
\epsfysize=3.5cm
\leavevmode
\epsfbox{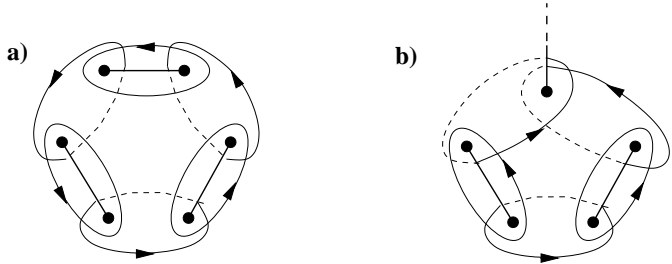}
\end{center}
\caption[]{\small A set of non-trivial $\gamma_{(i,i+1)}$ 1-cycles in two 
Riemann surfaces with even and odd $N$.}
\label{riemann}
\end{figure}

An interesting property to be used below is that there exists a simple set 
of 1-cycles, denoted $\gamma_{(i,i+1)}$, defined as closed paths 
surrounding (counterclockwise) the branch points $z_i$, $z_{i+1}$, and whose 
only non-vanishing intersection numbers are
\beqa
\gamma_{(i-1,i)}\cdot\gamma_{(i,i+1)}\, =\, 1
\label{interone}
\eeqa
with $i$ defined modulo $N$. Moreover, it is straightforward to use a 
deformation argument to show the homological relation
\beqa
\sum_{i=1}^N \, [\gamma_{(i,i+1)}]\, = \, 0
\label{relaone}
\eeqa
Examples of this set for even and odd $N$ are shown in figure 
\ref{riemann}.

We are interested in configurations of D6-branes wrapped on 3-cycles in 
$\IT^2\times \IT^2\times \IY$, given by a product of three 1-cycles, one 
per factor (As usual we skip the discussion of non-factorizable 3-cycles). 
Using 1-cycles of $\IY$ of the kind described above, each stack of 
D6$_a$-branes is labeled by its wrapping numbers in $\IT^2\times 
\IT^2$, namely $(n^1_a,m^1_a)$, $(n^2_a,m^2_a)$ and an index $i_a$ 
corresponding to its 1-homology class 
in $\IY$, $\gamma_{(i_a,i_a+1)}$. Denoting by $[\Pi]$ the 2-homology 
class of the 3-cycle in $\IT^2\times \IT^2$, the intersection number 
between the 3-cycles associated to the $a^{th}$ and $b^{th}$ stack is 
\beqa
I_{a,b} \, = \, [\Pi_a]\cdot [\Pi_b] \, \times \, (\delta_{i_b,i_a+1} - 
\delta_{i_b,i_a+1})
\eeqa
This number determines the number of arrows joining the nodes associated 
to $a$ and $b$ in the quiver diagram. 

The contributions to the RR tadpoles are easily computed by checking 
intersection numbers in the generic quiver, or by using 
homological relations like (\ref{relaone}) from deformation arguments.
For instance, denoting $[\Pi_i]$ the total 2-homology class in 
$\IT^2\times \IT^2$ associated with D6-branes on wrapped on the 1-cycle 
$\gamma_{(i,i+1)}$ in $\IY$, we may guarantee cancellation of RR tadpoles 
by requiring
\beqa
\sum_{i=1}^N\, [\Pi_i]\, = \, 0
\label{rrtadone}
\eeqa

Instead of building examples of this kind, let us introduce a further 
ingredient required to reproduce the Standard Model instersection numbers, 
namely the orientifold projection. We may for instance mod out by $\Omega$ 
times 
\beqa
z_1 \to {\ov z}_1 \quad ; \quad z_2 \to {\ov z}_2 \quad ; \quad z\to {\ov z}
\label{or}
\eeqa
on the complex coordinates $z_1$, $z_2$ of $\IT^2\times \IT^2$, and on 
(\ref{doublecover}). Such symmetry is present for onefolds $\IY$ with 
a set of branch points and cuts invariant with respect to the imaginary 
$z$ axis, as in figure \ref{riemann}. Also the 2-tori are restricted to be 
rectangular or tilted at a particular angle. In the latter case, we 
introduce (possibly fractional) effective wrapping numbers $(n,m)$ with 
$m$ integer or half-odd if $n$ is even or odd \cite{bkl}. 

We would like to describe an orientifold action on the homology classes 
of 3-cycle. Consider for concreteness the case of odd $N$, where 
due to $\IZ_2$ symmetry the unpaired branch cut must lie at imaginary 
$z$. Labelling such branch point as $i=(N+1)/2$, the orientifold action on 
3-cycles we are interested in is to change their labels as follows 
\footnote{Strictly speaking the orientifold action (\ref{or}) does not act 
like this on 1-cycles in $\IY$. On the other hand, it does act in this 
fashion on certain linear combinations of 1-cycles, which have the same 
intersection as the original ones. This implies that there exists an 
orientifold action, related to (\ref{or}) by a change of variables 
(symplectic transformation), which acts on 1-cycles as (\ref{swap}). To 
simplify the discussion we stick to the latter.}
\beqa
(n^1,m^1), (n^2,m^2), i \; \rightarrow (n^1,-m^1), (n^2,-m^2), -i 
\label{swap}
\eeqa

The contribution of the O6-plane to RR tadpoles can be computed using 
intersection numbers or deformation arguments. With this information it is 
now possible to look for configurations of D6-branes on 3-cycles which 
reproduce the Standard Model intersection numbers. However, instead of 
performing a brute force attempt, it will be useful to note that the same 
algebraic problem has already appeared in a different (but possibly 
related, see below) context. Such small detour will also make manifest our 
motivation to study the above threefolds and the  above set of 1-cycles in 
looking for the Standard Model intersection numbers. 

\medskip

One particular realization of the spectrum of just the Standard 
Model has been provided in \cite{cim} in (orientifolded) configurations of 
D5-branes wrapped on 2-cycles on $\IT^2\times \IT^2$ and sitting at a 
$\IC/\IZ_N$ singularity. In such models, stacks of D5-branes are labeled 
by an index $i$ defined modulo $N$, defining its Chan-Paton eigenvalue 
with respect to the $\IZ_N$ action. The chiral spectrum from strings 
stretched between the $a^{th}$ and $b^{th}$ stacks is given by
\beqa
I_{a,b} \, = \, [\Pi_a]\cdot [\Pi_b] \, \times \, (\delta_{i_b,i_a+1} - 
\delta_{i_b,i_a+1})
\label{numdfive}
\eeqa
where the last piece arises from the quiver structure of the $\IC/\IZ_N$ 
singularity.

The key point is that the numbers of bi-fundamental chiral fermions 
(\ref{numdfive}) in D5-brane models exactly coincides with the 
intersection numbers of 3-cycles in $\IT^2\times \IT^2\times \IY$. 
Moreover, the orientifold action on D5-brane stacks exactly coincides with 
the label exchange (\ref{swap}). Similarly, the translation of
constraints (\ref{rrtadone}) to D5-brane language reproduces the 
cancellation of twisted RR tadpoles for D5-branes at $\IC/\IZ_N$.

Hence the problem of finding explicit realizations of configurations of 
D6-branes wrapped on 3-cycles in $\IT^2\times \IT^2\times \IY$ is 
algebraically isomorphic to the problem of finding explicit models of 
D5-branes in $\IT^2\times \IT^2\times \IC/\IZ_N$. In \cite{cim} explicit 
models of D5-branes leading to the chiral spectrum of {\em just} the 
Standard Model were constructed. Using the above mapping, it is an 
straightforward exercise to describe explicit configurations of D6-branes 
on 3-cycles producing exactly such spectrum.

\begin{figure}
\begin{center}
\centering
\epsfysize=3.5cm
\leavevmode
\epsfbox{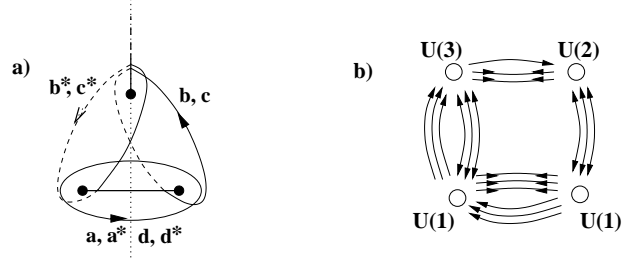}
\end{center}
\caption[]{\small Figure a) shows the structure of cycles used in the 
construction of the Standard Model example. Figure b) provides the quiver 
diagram for the resulting gauge theory, which indeed reproduces the 
Standard Model (plus right-handed neutrinos).}
\label{justsm}
\end{figure}

For instance, consider the configuration of D5-branes with wrapping 
numbers and Chan-Paton factors as in table 4 in \cite{cim} for the 
particular choice $n_a^1=n_d^1=1$, $\epsilon={\tilde \epsilon}= 
\epsilon_h=1$, $\beta^1=1/2$. This translates into a configuration of 
D6-branes in a geometry $\IT^2\times \IT^2\times \IY_3$, with 
$\IY_3$ having three branch points. The configuration contains D6-branes 
wrapped on the 3-cycles as follows \footnote{In order to obtain final 
intersection numbers in agreement with those in \cite{imr}, we have 
exchanged $d\leftrightarrow d^{*}$ with respect to \cite{cim}.} 
\beqa
a & 3\, (1,1/2) \times (3,-1/2)\times \gamma_{0,1} \nonumber \\
b & 2\, (2,0) \times (1,-1/2)\times \gamma_{1,2} \nonumber \\
c & 1\, (2,0) \times (0,1)\times \gamma_{1,2} \nonumber \\
d & 1\, (1,-3/2) \times (1,-1/2)\times \gamma_{0,1} \nonumber \\
h & 4\, (2,0) \times (2,0)\times \gamma_{0,1} 
\eeqa
and their orientifold images (denoted $a^*$, $b^*$, etc). The 
configuration is schematically shown in figure \ref{justsm}a).

The configuration can be checked to be free of tadpoles. Upon computation 
of intersection numbers, we obtain
\beqa
I_{ab}=1\quad & ; & \quad I_{ab^*}=2 \nonumber \\
I_{ac}=-3\quad & ; & \quad I_{ac^*}=-3 \nonumber \\
I_{bd}=0\quad & ; & \quad I_{bd^*}=-3 \nonumber \\
I_{cd}=-3\quad & ; & \quad I_{cd^*}=-3 
\label{interimr}
\eeqa
precisely reproducing the Standard Model intersection numbers (2.3) in 
\cite{imr}. It should be obvious how to map any other D5-brane model in 
\cite{cim} to a local configuration of D6-branes on 3-cycles with 
intersection numbers (\ref{imr}).

There is one last issue, crucial to indeed obtain the Standard Model 
spectrum. This is the constraint that a linear combination of $U(1)$'s 
remaining massless plays the role of hypercharge. Very remarkably it is 
possible to compute the $BF$ couplings in the setup of D6-branes on 
3-cycles and recover the same condition as in models of D5-branes. 

Hence the above setup provides a new realization of the Standard Model 
intersection numbers, now in terms of D6-branes wrapped on compact 
3-cycles in a non-compact threefold. Although the background geometry 
is non-supersymmetric, we find the result very satisfactory. The final 
spectrum is given in table~\ref{tabpssm}, taken from \cite{imr}.

\begin{table}[htb] \footnotesize
\renewcommand{\arraystretch}{1.25}
\begin{center}
\begin{tabular}{|c|c|c|c|c|c|c|c|}
\hline Intersection &
 Matter fields  &   &  $Q_a$  & $Q_b $ & $Q_c $ & $Q_d$  & Y \\
\hline\hline (ab) & $Q_L$ &  $(3,2)$ & 1  & -1 & 0 & 0 & 1/6 \\
\hline (ab*) & $q_L$   &  $2( 3,2)$ &  1  & 1  & 0  & 0  & 1/6 \\
\hline (ac) & $U_R$   &  $3( {\bar 3},1)$ &  -1  & 0  & 1  & 0 & -2/3 \\
\hline (ac*) & $D_R$   &  $3( {\bar 3},1)$ &  -1  & 0  & -1  & 0 & 1/3 \\
\hline (bd*) & $ L$    &  $3(1,2)$ &  0   & -1   & 0  & -1 & -1/2  \\
\hline (cd) & $E_R$   &  $3(1,1)$ &  0  & 0  & -1  & 1  & 1   \\
\hline (cd*) & $N_R$   &  $3(1,1)$ &  0  & 0  & 1  & 1  & 0 \\
\hline \end{tabular}
\end{center} \caption{ Standard model spectrum and $U(1)$ charges
\label{tabpssm} }
\end{table}               

\subsection{Mirror symmetry}

The above intricate agreement in the topological properties of D5-branes 
at $\IC/\IZ_N$ and D6-branes in $\IY$ demands a deeper explanation. We 
strongly suspect that this explanation arises from mirror symmetry between 
$\IC/\IZ_N$ and $\IY$, as we roughly explain. 

The mirror of $\IC/\IZ_N$ has been determined using the world-sheet 
techniques of \cite{hv} in \cite{vafatachyon}. It is described by (the 
infrared limit of) an $\NN=2$ supersymmetric two-dimensional 
Landu-Ginzburg model with superpotential
\beqa
w(x)\, =\, x^N \, + \, f(x)
\label{onefold}
\eeqa
where $x$ is a chiral multiplet, and $f(x)$ contains lower order terms 
which deform the structure of the vacua (and are related to the amount of 
condensation of closed string twisted tachyons). Its form will not be too 
important for most of our discussion.

From the geometric viewpoint, the mirror is a non-compact one complex 
dimensional space $\ItY$ described as the hypersurface 
(\ref{onefold}) in the $\IC^2$ parametrized by $w$, $x$. It is an $N$-fold 
cover of the $w$ plane, with $N-1$ critical points $w_{*,i}$ at which 
$dw/dx=0$. For each value of $w$ there exist $N$ values of $x$ which are 
pre-images of the map $w(x)$ and so project down to the same point. These 
values of $x$ are distinct, except at the cricical values $w_{*,i}$, where 
two pre-images coincide \footnote{Such values of $x$ correspond to the 
(generically massive) vacua of the two-dimensional world-sheet theory.}. 
Hence as $w$ varies (\ref{onefold}) describes $N$ generically disconneted
branches, a pair of which touch at each critical point (which pairs join 
at which critical point depends on the structure of $w$). Asymptotically 
at large $w$ the $k^{th}$ branch tends to $x=w^{1/N}e^{2\pi i\, k/N}$.

\begin{figure}
\begin{center}
\centering
\epsfysize=3.5cm
\leavevmode
\epsfbox{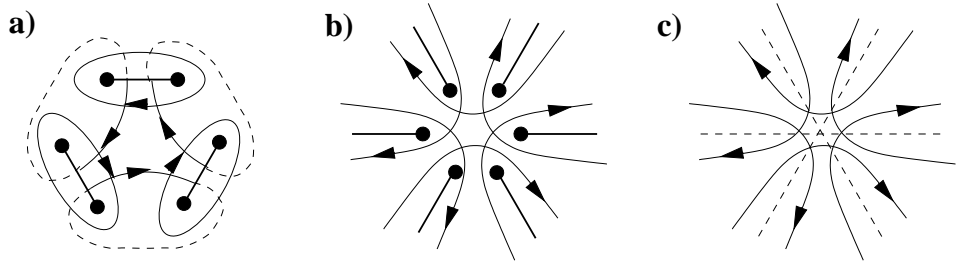}
\end{center}
\caption[]{\small Schematic depiction of the relation between 1-cycles in 
the geometries $\IY$, $\ItY$. Starting from 1-cycles in $\IY$, one may 
deform the branch cuts to reach b). Since the 1-cycles in b) do not cross 
any branch cuts, the double cover structure is not too relevant, and the 
1-cycle geometry reproduces that of 1-cycles in $\ItY$, figure c.}
\label{mirrory}
\end{figure}

The structure of the 1-cycles on which D-branes can wrap in $\ItY$ 
has been determined in \cite{hiv} (section 5.2). They correspond to real 
curves coming in from infinity along $x=w^{1/N}e^{2\pi i\, k/N}$, 
reaching a critical value where the $k^{th}$ and $k'^{th}$ branches touch, 
and going off to infinity along $x=w^{1/N}e^{2\pi i\, k'/N}$. Such 
1-cycle, denoted $\gamma_{(k,k')}$, has a projection in the $x$-plane 
which looks 
like a wedge of angle $2\pi (k-k')/N$. Roughly speaking \footnote{Which 
1-cycles indeed exist depends on the critical point structure and hence 
on $W$. We assume a suitable choice so that the following basis exists.} a 
basis of 1-cycles is provided by ${\tilde \gamma}_{(i,i+1)}$ with 
$i=1,\ldots, N$ with the homological relation $\sum_{i=1}^N 
{\tilde \gamma}_{(i,i+1)}=0$, see figure \ref{mirrory}c.

The intersection numbers of these 1-cycles \footnote{Since the 1-cycles 
are non-compact, the intersection numbers are more properly defined as the 
number of solitons between the two corresponding two-dimensional vacua of 
the LG theory.}, has been computed in \cite{hiv}, sections 2.4.1 and 5.2 
to be
\beqa
\gamma_{i-1,i}\cdot\gamma_{i,i+1}\, = \, 1
\label{intone}
\eeqa
This precisely reproduces the quiver diagram of the $\IC/\IZ_N$ orbifold, 
as expected from mirror symmetry.

There is a remarkable analogy between the structure of 1-cycles in the 
geometry $\ItY$ and our onefolds $\IY$. This relation seems to go beyond 
mere topology, and hold even at the level of the holomorphic structure, 
since the geometries may be related by a transformation preserving it, as 
we show in figure \ref{mirrory}. Hence, we strongly suspect that there is 
a weak equivalence relation between the geometries $\ItY$, $\IY$, in the 
sense of \cite{hv}, which allows to consider $\IY$ as the mirror of 
$\IC/\IZ_N$. This relation would therefore explain the close relation of 
our D6-brane models in this section to models of D5-branes.

\subsection{Relation to models on $\IT^6$ with infinite throats}

It is interesting to compare the configurations with Standard Model 
intersection numbers, realized in terms of D6-branes on $\IT^2\times 
\IT^2 \times \IT^2$ vs on $\IT^2\times \IT^2\times \IY$. One main 
difference is that the latter are non-compact geometries; hence (once 
plugged into a global compact model) allow to lower the string scale 
while maintaining a large Planck scale, and preventing the appearance of 
too light Kaluza-Klein gauge bosons. Interestingly  enough, in the case 
discussed above of $\IY$ with three branch points, denote it $\IY_3$, it 
may be regarded as a two-torus with an infinite throat rendering it 
non-compact. From this viewpoint, models constructed by gluing 
$\IT^2\times \IT^2\times \IY_3$ to a global compact large space
may be regarded as models of D6-branes wrapped on a local region which is 
almost $\IT^6$, but which contains a throat connecting it with the rest 
of (an arbitrarily large) geometry. The models therefore provide an explicit 
realization of the mechanism proposed in \cite{afiru} to solve the low 
string scale problem.

In fact, this suggests an alternative way of constructing Standard 
Model local configurations of D6-branes on 3-cycles in $\IT^2\times 
\IT^2\times \IY_3$. Namely, one may take any model in table 2 in 
\cite{imr}, and map the homology class of the 1-cycles in one of the 
tilted 2-tori (by \cite{bkl} there must be at least one) to homology 
1-cycles in $\IY_3$, via the rule
\beqa
[a] \to [\gamma_{(1,2)}] \quad ; \quad [b]\to [\gamma_{(1,2)}] - 
[\gamma_{2,0}]
\eeqa
This maps consistently the orientifold action (\ref{or}) on 1-cycles in 
$\IY_3$ with with the orientifold action $[b]\to -[b]$, $[a]\to [a]-[b]$ 
for the tilted 2-tori in \cite{imr}; the mapping also preserves 
intersection numbers (modulo an irrelevant overall sign). Using this 
mapping the models of D6-branes on $\IT^6$ can be translated to models of 
D6-branes on $\IT^2\times \IT^2\times \IY_3$, preserving the physical 
properties of the corresponding gauge theory: for instance, the chiral 
spectrum, or more remarkably the $BF$ couplings of $U(1)$ generators to 
RR fields. This allows the construction of large classes of models of 
D6-branes on 3-cycles in $\IT^2\times \IT^2\times \IY_3$ with just the 
Standard Model spectrum; we leave the details to the reader.

Hence the above trick provides a systematic way of starting from a purely 
toroidal $\IT^6$ model and mapping it to a model in a geometry with is 
$\IT^6$ with an infinite throat glued to it. More generally, one may 
consider threefolds which are products of non-compact Riemann surfaces. 
This is advantageous since it introduce more non-compact dimensions, and 
facilitates the task of generating large Planck masses from low string 
scales. The models may easily be constructed by starting with $\IT^6$ 
models and mapping the 1-homology classes on the $\IT^2$ factors to 
homology classes in $\IY_3$ factors. At the price of giving up 
supersymmetry in the closed string sector (which may be a moderate one, 
given that we may lower the string scale), one may construct large classes 
of models which are toroidal in the neighbourhood of the D6-branes, but 
contain infinite throats only felt by closed string sector modes.

\section{Conclusions}
\label{conclusions}

Our purpose in the present paper has been to point out the existence of 
large classes of relatively simple non-compact Calabi-Yau threefolds with 
compact 3-cycles intersecting at points. They may be easily used to 
construct local model of four-dimensional chiral gauge theory sectors, by 
wrapping D6-branes on such 3-cycles. We have also provided the basic rules 
to compute their intersection numbers and contributions to RR tadpoles, 
and have illustrated diverse model building possibilities with explicit 
examples, some of them with phenomenologically very appealing spectrum. 
In fact, by relaxing the constraint of having a supersymmetry preserving 
geometry, we have suceeded in constructing (moreover, in giving precise 
rules to construct large classes of) explicit models with {\em just} the 
Standar Model spectrum.

We would like to emphasize once more that, even though the threefolds we 
have centered on first appeared in the context of mirror symmetry, the 
manifolds can be constructed without help from mirror symmetry. In fact, 
the strategy should be to construct threefolds with interesting structure 
of intersecting 3-cycles, and interesting configurations of D6-branes 
wrapped on them, regardless of their mirror symmetry properties.

There are several directions that should be explored in the present setup. 
A more complete classification of the kind of threefolds with compact 
3-cycles would be desirable. Even in the concrete class studied in this 
paper, a better control over the sets of $(p,q)$ degenerations that are 
allowed consistently with the Calabi-Yau condition would be desirable (or 
even if one relaxes beyond supersymmetric geometries). Also, given a 
particular geometry, it would be useful to find a systematic and efficient 
way of describing a basic set of compact and non-compact 3-cycles 
generating the complete (compact and non-compact) 3-homology, in order to 
allow a more systematic search for consistent D6-brane configurations, and 
a systematic computation of their generic spectra. Such formal 
developments seem quite crucial in providing a general view on the class 
of gauge theories which may be engineered in this setup, and to search 
for particular solutions with specific spectra (for instance, the 
Standard Model intersection numbers in \cite{imr}). We hope to report on 
these ideas in the future.

A final open avenue which we would like to mention is the M-theory lift of 
configurations of this kind preserving $\NN=1$ supersymmetry. As discussed 
in e.g. \cite{gomis}, they should correspond to purely geometrical 
configurations involving a non-compact 7-manifold admitting a $G_2$ 
holonomy metric. The possibility of obtaining non-abelian gauge 
interactions and chiral fermions from such $G_2$ configurations had been 
noticed in \cite{aw}, and proposed to be used for phenomenological 
purposes. Unfortunately, compact $G_2$ manifolds are 
difficult to construct, beyond the implicit definition as lifts of IIA 
configurations with O6-planes and D6-branes \cite{km,susy,susy2}. On the 
other hand, the only known non-compact $G_2$ manifolds lead to 
non-localized gauge interactions, corresponding to A-D-E singularities 
fibered over non-compact 3-manifolds \cite{aw}.

The kind of type IIA configurations we have studied, however, would lead 
to completely four-dimensional gauge interactions and chiral fermions, 
despite the fact that the complete $G_2$ manifold is non-compact. Some of 
the type IIA configurations we have studied, like that in section 
\ref{trinification}, seem symmetric enough to have a relatively simple 
lift. Although their $G_2$ metrics are presumably beyond present 
techniques (even the metric for the base Calabi-Yau is not known), the 
topology of the seven-manifold is likely to be computable and lead to a 
simple answer. Such a result would be interesting to e.g. reinterpret 
tadpole cancellation constraints in terms of geometry (a related 
discussion is in \cite{anomgtwo}); but more importantly would provide the 
first realization of four-dimensional chiral gauge theories genuinely from 
M-theory on $G_2$ manifolds.

\centerline{\bf Acknowledgements}

I thank L.~E.~Ib\'a\~nez for useful discussions. I also thank
M.~Gonz\'alez for kind support and encouragement.

\end{document}